\documentclass[12pt]{article}
\usepackage{epsfig}
\usepackage{amsfonts}
\usepackage{amscd}
\usepackage{latexsym}
\usepackage{amsmath,amssymb,amsthm}
\usepackage{verbatim}
\usepackage{setspace}
\usepackage{color}
\usepackage{cite}
\usepackage{graphicx}
\usepackage{mathtools}
\usepackage[all]{xy}
\usepackage{tikz}
\usetikzlibrary{patterns}

\usepackage[textheight=9in, textwidth=6.5in, letterpaper]{geometry}

\usepackage{color}   
\usepackage{hyperref}
\hypersetup{
    colorlinks=true,  
    linktoc=all,     
    linkcolor=black,  
    citecolor=black,
    filecolor=black,
    urlcolor=black,
}

\numberwithin{equation}{section}

\def\p{\partial}

\def\<{\langle}
\def\>{\rangle}

\def\cO{\mathcal{O}}

\def\be{\begin{equation}}
\def\ee{\end{equation}}
\def\beq{\be\begin{array}{c}}
\def\eeq{\end{array}\ee}
\def\bes{\be\begin{split}}
\def\ees{\end{split} \ee}
\def\bs{\begin{split}}
\def\es{\end{split} }

\def\a{{\alpha}}
\def\g{{ \gamma}}

\def\e{{\epsilon}}

\def\G{\Gamma}

   \makeatletter
  \let\over=\@@over \let\overwithdelims=\@@overwithdelims
  \let\atop=\@@atop \let\atopwithdelims=\@@atopwithdelims
  \let\above=\@@above \let\abovewithdelims=\@@abovewithdelims
\renewcommand\section{\@startsection {section}{1}{\z@}%
                                   {-3.5ex \@plus -1ex \@minus -.2ex}
                                   {2.3ex \@plus.2ex}%
                                   {\normalfont\large\bfseries}}

\renewcommand\subsection{\@startsection{subsection}{2}{\z@}%
                                     {-3.25ex\@plus -1ex \@minus -.2ex}%
                                     {1.5ex \@plus .2ex}%
                                     {\normalfont\bfseries}}

\begin{document}
\begin{titlepage}
\unitlength = 1mm

\vskip 1cm
\begin{center}

{ \LARGE {\textsc{Tropical mirror for toric surfaces}}} \\

\vspace{1cm}

Andrey Losev\\
\vspace{0.3cm}
{\it Wu Wen-Tsun Key Lab of Mathematics, Chinese Academy of Sciences, USTC,
No.96, JinZhai Road Baohe District, Hefei, Anhui, 230026, P.R.China \\
\vspace{0.3cm}
National Research University Higher School of Economics, Laboratory of Mirror Symmetry, NRU HSE, 6 Usacheva str., Moscow, Russia, 119048\\
}

\vspace{1cm}
 Vyacheslav Lysov\\
 \vspace{0.3cm}
{\it Okinawa Institute of Science and Technology,\\
 1919-1 Tancha, Onna-son, Okinawa 904-0495, Japan}
\vspace{1cm}

\begin{abstract}
We describe the tropical mirror for  complex toric surfaces.   In particular we provide an explicit expression for the mirror states and show that they can be written 
in enumerative form.   Their holomorphic germs give an explicit form of   good section for Landau-Ginzburg-Saito theory. We use an explicit form of holomorphic germs to derive the divisor relation for tropical Gromov-Witten invariants.  
We interpret the deformation of the theory by a point observable as a blow up of a point on  the toric surface. We describe the implication of such interpretation for the tropical  Gromov-Witten invariants.
\end{abstract}

\vspace{1.0cm}

\end{center}

\end{titlepage}

\pagestyle{empty}
\pagestyle{plain}

\pagenumbering{arabic}

\tableofcontents

\section{Introduction}

Tropical mirror symmetry has all features of the mirror symmetry while providing a much simpler description for most of them.  In particular,   holomorphic curves become graphs and topological string theory becomes topological quantum mechanics.  In our paper we argue that the same level of simplification holds for the mirror of the evaluation observables. 

The conventional mirror symmetry \cite{vafa2003mirror} focuses on the superpotential and the choice of special coordinates on its space of deformations.  The  choice of   special coordinates is encoded as a solution to a certain dynamical system (starting from 
pioneering work by K. Saito  \cite{saito1983period}), which can be phrased as flatness and torsionless  condition for some connection. The Christoffel symbols for this  connection can be encoded as a contact terms determined by  K. Saito's good section.  Using this method, in order to evaluate the $n$-point invariant we need to differentiate the 
3-point invariant, given by the residue formula, $n-3$ times with respect to the  special coordinates. 

In  our approach to tropical mirror we focus on observables rather than the superpotential. The contact terms  naturally emerge as  distinguished deformation of the mirror states in topological quantum mechanics. 
Such distinguished deformations for polynomial superpotentials were constructed in \cite{Losev:1998dv}, for a holomorphic germination of  harmonic form states. 
Hence, we can immediately describe the tropical good section for Landau-Ginzburg-Saito theory. Moreover, we can directly evaluate  the correlation functions   using the mirror states for the evaluation observables.    

Given various simplifications of the mirror map in the tropical approach we can expect  that the mirror states could also have an explicit  description. In our work \cite{Losev:2022tzr,Losev:2023bhj}  we provided an integral representation for the mirror states. Moreover, for the case of $\mathbb{P}^1$ the integrals evaluate into the  indicator  functions. However, the simplicity of the answers might be the feature of simplest example, hence we evaluated the mirror states for the observables on a  2-dimensional toric surfaces.  

In this paper we  show   that the mirror states  can be written using the indicator  functions on  cones, which are standard objects in toric (algebraic) geometry  \cite{oda1988convex,telen2022introduction}.  Moreover, we showed that the sum over the indicator  functions can be rewritten as a weighted sum over  intersection points of particular graphs.  Similar sums were introduced by Mikhalkin  \cite{Mikhalkin2004,Mikh,mikhalkin2009tropical}  to define the intersection number for tropical curves. 

Given an explicit form of the holomorphic germs we can use the Landau-Ginzburg-Saito theory  to check one of the universal relations for the Gromov-Witten invariants:  the divisor relation.  In present paper we derive  the   divisor relation  from the recursion formula for the correlation functions in Landau-Ginzburg-Saito theory.      In particular,  we use our expression for the holomorphic germs of the   hyperplane observables  to show that they change moduli of the superpotential, while preserving the topology of the toric space.  

An explicit form of holomorphic germs allows us to give an explicit form of the tropical good section for the toric surfaces.  Note that  already for polynomial superpotentials with more than one variable it is possible to have more than one good sections, so it might be hard to choose one, relevant for the mirror symmetry. Our construction of the good section uses the holomorphic germs of the mirror states. 

The last but not least application of the mirror states in explicit form allows us to describe a (novel?) relation between the Gromov-Witten invariants on $\mathbb{P}^2$ and the $Bl_0(\mathbb{P}^2)$. We call it the ``cutting corners" relation. The relation is similar to the divisor relation. The $(n+1)$-point function   with a point observable on $\mathbb{P}^2$  is related to the $n$-point function on  $Bl_0(\mathbb{P}^2)$. 

 The structure of our paper is as follows:   In section 2 we briefly review the relevant information on the geometry of smooth complex toric surfaces. In section 3 we briefly review the tropical mirror map and  describe the mirror states and holomorphic germs of the observables on toric surface. In section 4 we derive the divisor relation from the recursion formula in Landau-Ginzburg-Saito theory and our explicit expression for the holomorphic germ of the hypersurface observable.
 In section 5 we describe the mirror for the several simples toric surfaces $\mathbb{P}^2, \mathbb{P}^1\times \mathbb{P}^1$ and $Bl_0(\mathbb{P}^2)$.  In section 6 we present  the cutting corners procedure for the $\mathbb{P}^2$  and 
 formulate the related open questions and conjectures.

\section{Geometry of toric surfaces}

In this section we will briefly review the geometry for 2-dimensional toric varieties, equivalently complex toric surfaces. 

\subsection {Projective toric surface}

 Toric    surface $X$ is a compactification of ${\mathbb{C^\ast}}^2$. We can represent $\mathbb{C}^{\ast 2} = \mathbb{R}^2 \times \mathbb{T}^2$ in the form of  the radial part $\mathbb{R}^2$, equipped  with standard coordinates $r^i, \; i=1,2$ and angular part,  $2$-dimensional torus  $\mathbb{T}^2 = S^1\times S^1$, with standard angular  coordinates $\phi^i$. Equivalently, we can say that the  $\mathbb{C}^{\ast 2}$ is a trivial $2$-dimensional  toric fibration over  $\mathbb{R}^2$.  We describe the compactification of ${\mathbb{C^\ast}}^2$  using the fibration data. 

\begin{itemize}

\item  The radial part $\mathbb{R}^2$  is compactified by  {\it convex rational  polytope}.  We will  describe a polytope by a collection of {\it supporting hyperplanes}.
 
\item  Each hyperplane is given in terms of the inside-pointing $2$-dimensional   (normal) vector $\vec{b}$ with components $b^i, i=1,2$.   For rational polytope each   vector has integer components i.e. $b^i \in \mathbb{Z}$.    For toric space $X$ we will denote the set of  corresponding  vectors by $B_X$.

\item In order to get a compactification of a complex manifold, we require that one of the circles $S^1 \subset \mathbb{T}^2$  inside the  toric fibration  shrinks to zero when we approach each of the compactifying hypersurfaces. The choice of a circle is given by a class in $\pi_1( \mathbb{T}^2) $ defined by a  normal vector $\vec{b}$ of the hyperplane . 
\end{itemize}

In toric geometry  \cite{oda1988convex,telen2022introduction}  the collection of normal vectors  $B_X$  define a  {\it fan} of $X$, hence we will adopt this notation for $B_X$.  Let us order vectors $\vec{b} \in B_X$ in counterclockwise order on $\mathbb{R}^2$. 
The consecutive pairs form cones of a fan  for $X$. A 2-dimensional  cone, formed by pair of vectors $\vec{b}_1$ and $\vec{b}_2$ is 
\be
\hbox{Cone} (\vec{b}_1, \vec{b}_2) = \{ \vec{b}_1 \;t_1 + \vec{b}_2\;t_2\;| \;t_1, t_2 \in \mathbb{R}^{\geq 0}\} \subset \mathbb{R}^2.
\ee 

 We will restrict our consideration to the smooth toric surfaces.  The fan for a smooth toric surface consists of smooth cones. Smoothness of  Cone$(\vec{b}, \vec{c})$  requires that the generating vectors form a basis in $\mathbb{Z}^2$, what is equivalent to 
 \be
 \det(\vec{b}, \vec{c}) = \pm 1.
 \ee
It is convenient to   introduce   a cross product  for two vectors,  so that 
\be
 \det(\vec{b}, \vec{c}) =b^1 c^2 - b^2 c^1  =  \vec{b}\times \vec{c}. 
\ee
Note that the sign of the  cross product $ \vec{b}\times \vec{c}$ is determined  by  the relative orientation of  the two vectors. The sign is positive  if we can rotate (angle less than $\pi$) from $\vec{b}$ to $\vec{c}$ in counterclockwise direction
and negative otherwise.

\subsection{Rays and stars}

By   construction a genus-0 smooth tropical curve is  an embedding of a 3-valent  tree  into $\mathbb{R}^2$ by a piece-wise linear map. For more details see Mikhalkin  \cite{Mikhalkin2004,Mikh,mikhalkin2009tropical}. The leaves of a tree
map  to infinite rays  along the  normal vectors of compactifying polytope. Moreover, each  tropical curve requires a balance condition: the sum of all vectors on the leaves equals to zero.  Below there are four examples of tropical curves, drawn in corresponding compactifying  polytopes.
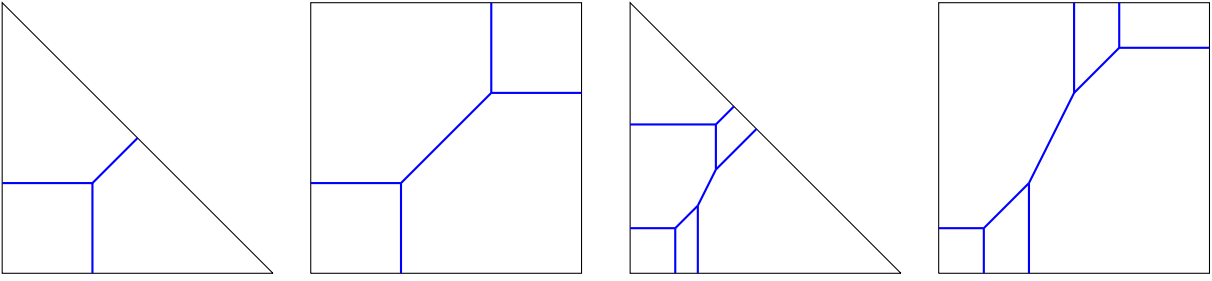
\newline\newline
\begin{tikzpicture}[scale=0.6]
\draw(4,-2)-- (-2, -2) -- (-2, 4)--(4,-2);
\draw[color=blue, thick] (-2, 0) -- (0, 0)-- (0,-2);
\draw[color=blue, thick](0, 0)-- (1,1);
\end{tikzpicture}\;\;\;
\begin{tikzpicture}[scale=0.6]
\draw(-2,-2)-- (4, -2) -- (4, 4)--(-2,4)--(-2,-2);
\draw[color=blue, thick] (-2, 0) -- (0, 0)-- (0,-2);
\draw[color=blue, thick](0, 0)-- (2,2);
\draw[color=blue, thick](2, 2)-- (2,4);
\draw[color=blue, thick](2, 2)-- (4,2);
\end{tikzpicture} \;\;\;
\begin{tikzpicture}[scale=0.6]
\draw(4,-2)-- (-2, -2) -- (-2, 4)--(4,-2);
\draw[color=blue, thick] (-2, -1) -- (-1, -1);
\draw[color=blue, thick]  (-1, -1)-- (-1,-2);
\draw[color=blue, thick](-1, -1)-- (-0.5,-0.5);
\draw[color=blue, thick](-0.5,-0.5)--(-.5,-2);
\draw[color=blue, thick](-0.5, -0.5)-- (-0.5+0.4,-0.5+0.8);
\draw[color=blue, thick](-0.5+0.4,-0.5+0.8)--(-0.1,-0.5+0.8+1);
\draw[color=blue,  thick] (-0.5+0.4,-0.5+0.8)-- (-0.5+0.4+0.9,-0.5+0.8+0.9);
\draw[color=blue, thick](-2,-0.5+0.8+1)--(-0.1,-0.5+0.8+1);
\draw[color=blue, thick](-0.1,-0.5+0.8+1)--(-0.1+0.4,-0.5+0.8+1+0.4);
\end{tikzpicture}\;\;\;
\begin{tikzpicture}[scale=0.6]
\draw(-2,-2)-- (4, -2) -- (4, 4)--(-2,4)--(-2,-2);
\draw[color=blue, thick] (-2, -1) -- (-1, -1);
\draw[color=blue, thick]  (-1, -1)-- (-1,-2);
\draw[color=blue, thick](-1, -1)-- (0,0);
\draw[color=blue, thick](0,0)--(0,-2);
\draw[color=blue, thick](0,0)-- (1,2)--(1,4);
\draw[color=blue, thick](1,2)--(2,3)--(2,4);
\draw[color=blue, thick](2,3) -- (4,3);
\end{tikzpicture} 
\newline\newline
 For any tropical curve we can construct its  (maximally) degenerate version by shrinking  the images of all internal edges of a tree to zero size. The resulting tropical curve will have a star shape.   Below we provide the results of shrinking procedure for the four curves above.
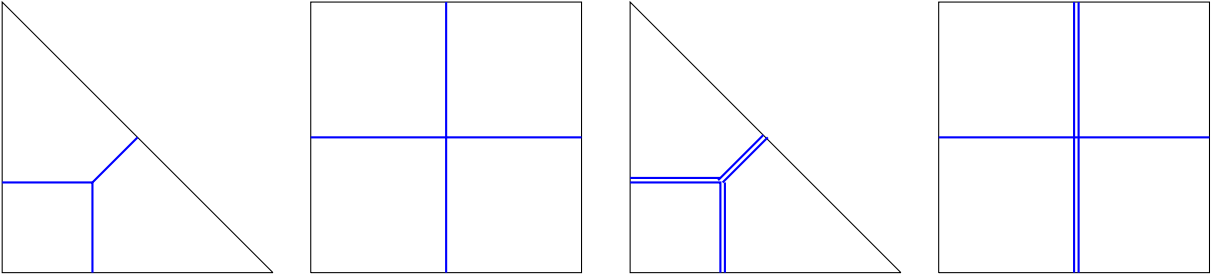
\newline\newline
\begin{tikzpicture}[scale=0.6]
\draw(4,-2)-- (-2, -2) -- (-2, 4)--(4,-2);
\draw[color=blue, thick] (-2, 0) -- (0, 0)-- (0,-2);
\draw[color=blue, thick](0, 0)-- (1,1);
\end{tikzpicture}\;\;\;
\begin{tikzpicture}[scale=0.6]
\draw(-2,-2)-- (4, -2) -- (4, 4)--(-2,4)--(-2,-2);
\draw[color=blue, thick] (-2, 1) -- (1, 1)-- (1,-2);
\draw[color=blue, thick](1, 1)-- (1,4);
\draw[color=blue, thick](1, 1)-- (4,1);
\end{tikzpicture} \;\;\;
\begin{tikzpicture}[scale=0.6]
\draw(4,-2)-- (-2, -2) -- (-2, 4)--(4,-2);
\draw[color=blue, thick] (-2, 0) -- (0, 0)-- (0,-2);
\draw[color=blue, thick] (-2, 0.1) -- (0, 0.1);
\draw[color=blue, thick]  (0.1, 0)-- (0.1,-2);
\draw[color=blue, thick](0.05, 0)-- (1.05,1);
\draw[color=blue, thick](-0.05, 0.05)-- (0.95,1.05);
\end{tikzpicture}\;\;\;
\begin{tikzpicture}[scale=0.6]
\draw(-2,-2)-- (4, -2) -- (4, 4)--(-2,4)--(-2,-2);
\draw[color=blue, thick](1, -2)-- (1,4);
\draw[color=blue, thick](1.1, -2)-- (1.1,4);
\draw[color=blue, thick](-2, 1)-- (4,1);
\end{tikzpicture}
\newline\newline
{\bf Definition}: Given a point $\vec{\rho}$  and a vector $\vec{l} $ we define a  ray $R_{l, \rho}$,  starting at $\rho$ and directed  along $\vec{l}$, i.e.
\be
R_{l, \rho} =\vec{\rho}+ \mathbb{R}^+ \vec{l} = \left\{( \rho^1+t\; l^1,  \rho^2+t\;l^2 )\in \mathbb{R}^2\;|\; t\in \mathbb{R}^+\right\} .
\ee 
A ray $R_{l, \rho}$ describes a holomorphic disc with Poincare-dual form 
\be
\g_{R_{l, \rho} }=  \frac{1}{(2\pi)^2}\int_{S^1} \int^\infty_0 \; \delta^2(\vec{r}-\vec{\rho}- \vec{l}t) (dr^1-l^1 dt) (d\phi^1-l^1 d\varphi)(dr^2-l^2dt)(d\phi^2-l^2 d\varphi),
\ee
which can be simplified  into
\be\label{Poincare_dual_ray}
\g_{R_{l, \rho} }= \frac{1}{2\pi}(\vec{l}\times d\vec{r})(\vec{l}\times d\vec{\phi})  \int^\infty_0  dt \; \delta^2(\vec{r}-\vec{\rho}- \vec{l}\;t).
\ee
{\bf Definition}: A {\it   star }  $S_\rho$  on complex  toric  surface  $X$   is  the union  of   rays from common end point  $\vec{l}$  from $\vec{\rho}$  
\be
S_{\rho} = \bigcup\limits_{\vec{l}\in S_\rho} R_{l, \rho}\;\;\;,
\ee
 such that  each  vector of s star   $\vec{l} =-\vec{b}$ for some   $\vec{b}  \in B_X$ and the sum of all   vectors  equals to zero. 
\be\label{balanced_star}
\sum_{\vec{l}\in S_\rho}  \vec{l} = 0.
\ee
The  equality (\ref{balanced_star}) is known as the {\it balancing condition}.   Note that there could be multiple rays in the same direction as depicted in examples above. The Poincare-dual of a star  is  a sum of the Poincare-duals  of all its rays, i.e. 
\be
\g_{S_\rho} = \sum_{\vec{l}\in S_\rho} \g_{R_{l, \rho}} \;.
\ee

\subsection{Intersection of rays and stars}

 Pair of rays   $R_{l, \rho} $ and $R_{n, 0}$ on a plane $\mathbb{R}^2$ may intersect at most at one point. We can  express the number of intersection points using Poincare-duals for the rays 
\be\label{int_rays_cone}
\begin{split}
R_{l, \rho} \cdot_{\mathbb{R}} R_{n, 0} & = \int_{\mathbb{R}^2}   (\vec{l}\times d\vec{r})  \int^\infty_0  dt_1\; \delta (\vec{r}-\vec{\rho}- \vec{l}\; t_1)  \wedge(\vec{n}\times d\vec{r})  \int^\infty_0 dt_2\; \delta (\vec{r}- \vec{n}\;t_2) \\
&=(\vec{l}\times \vec{n}) \int_{(\mathbb{R}^{+})^2} dt_1\;dt_2\; \delta (\vec{\rho}+ \vec{l}\;t_2-\vec{n}\;t_1)\; = \frac{(\vec{l}\times \vec{n})}{|\vec{l}\times \vec{n}|} \chi_{-\vec{l}, \vec{n}} (\vec{\rho}\;)\;,
\end{split}
\ee
where  we introduced an indicator  function, which equals to one inside the cone and to zero outside
\be
\chi_{\vec{l}_1, \vec{l}_2} (\vec{\rho}\;)  = \left\{
\begin{array}{ll}
1, & \rho \in \hbox{Cone} (\vec{l_1}, \vec{l_2}) \\
0,  &\rho \notin \hbox{Cone} (\vec{l_1}, \vec{l_2}).
\end{array}
 \right.
\ee
The denominator in (\ref{int_rays_cone}) is due to the Jacobian for the change of variables in the integral representation of the indicator function
\be\label{int_rep_theta_cone}
 \chi_{ \vec{l}_1, \vec{l}_2}(\vec{r}) =\int_{\hbox{Cone} (\vec{l}_1, \vec{l}_2)}  d^2 \vec{s}\;  \delta(\vec{r} - \vec{s})   =    |\vec{l}_1\times \vec{l}_2| \; \int^\infty_0 \int^\infty_0  dt_1 dt_2\;  \delta (\vec{r}-\vec{l}_1\; t_1-\vec{l}_2\;t_2 ).
\ee
The sign  factor for the intersection number (\ref{int_rays_cone}) is common feature for the intersection of real  cycles in real spaces. 

Our  formula (\ref{int_rays_cone})  tells us that the question of intersection for two rays is the same as a question whether  $\vec{\rho}$  belongs to a cone Cone$(\vec{n}, -\vec{l})$. Below we present the graphical proof of the relation $(\ref{int_rays_cone})$.
\newline\newline
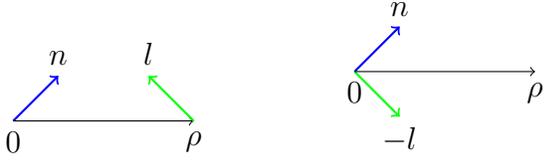
\begin{tikzpicture}[scale=0.6]
\draw[color=blue, thick, ->](0, 0)-- (1,1);
\node  at (1,1) [anchor=south]{$n$};
\draw[ ->](0, 0)-- (4,0);
\node  at (4,0) [anchor=north]{$\rho$};
\node  at (0,0) [anchor=north]{$0$};
\draw[color=green, thick, ->](4, 0)-- (3,1);
\node  at (3,1) [anchor=south]{$l$};
\end{tikzpicture}\qquad\qquad
\begin{tikzpicture}[scale=0.6]
\draw[color=blue, thick, ->](0, 0)-- (1,1);
\draw[color=green, thick, ->](0, 0)-- (1,-1);
\node  at (1,1) [anchor=south]{$n$};
\node  at (1,-1) [anchor=north]{$-l$};
\draw[ ->](0, 0)-- (4,0);
\node  at (4,0) [anchor=north]{$\rho$};
\node  at (0,0) [anchor=north]{$0$};
\end{tikzpicture}
\newline\newline
An intersection of two holomorphic discs, corresponding to the rays   in tropical geometry,   can be  expressed using the Poincare-dual  forms (\ref{Poincare_dual_ray}) via
\be\label{ray_intersect}
\begin{split}
R_{l,\rho} \cdot  R_{n, 0} &=   \int_{\mathbb{C}^{\ast 2}}  \g_{R_{l, \rho}} \wedge \g_{R_{n, 0}} \\
& = (\vec{l}\times \vec{n})^2 \int_{(\mathbb{R}^{+})^{2}} dt_1\;dt_2\; \delta (\vec{\rho}+ \vec{l}\;t_1-\vec{n}\; t_2)\; =|\vec{l}\times \vec{n}| \; \chi_{-\vec{l}, \vec{n}} (\vec{\rho}\;) .
\end{split}
\ee
An intersection of two tropical curves corresponding to two  stars, one at origin, other at $\vec{\rho}$  is given by the the intersections of all possible pairs of their rays 
\be\label{int_stars_cone}
S_\rho \cdot S_0' = \sum_{\vec{l}\in S,\;\; \vec{l}' \in S'}  R_{l,\rho} \cdot  R_{l', 0}    =   \sum_{\vec{l}\in S,\;\; \vec{l}' \in S'}  |\vec{l}\times\vec{l}'| \; \chi_{\vec{l}, -\vec{l}'} (\vec{\rho}\;).
\ee
{\bf Proposition}:  The intersection number  (\ref{int_stars_cone}) is  independent of the relative position $\vec{\rho}$.
\newline\newline
{\bf Proof}:  The sum (\ref{int_stars_cone}), defining the intersection number  changes when $\vec{\rho}$ crosses a boundary of a cone. Let us  assume that change in relative position  $\vec{\rho} \to {\vec{\rho}\;}'$ intersects a  single  ray along $\vec{l}_0'$ as shown on the picture below.
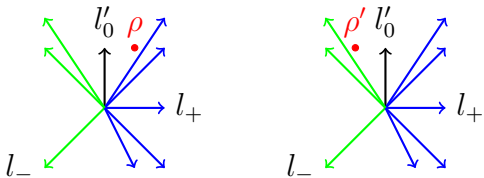
\newline\newline
\begin{tikzpicture}[scale=0.4]
\draw[color=blue, thick,->](0, 0)-- (2, 0);
\draw[color=black, thick, ->] (0, 0)-- (0,2);
\draw[color=green, thick, ->](0, 0)-- (-2,-2);
\draw[color=blue, thick, ->](0, 0)-- (2,2);
\draw[color=blue, thick, ->](0, 0)-- (2,3);
\draw[color=blue, thick, ->](0, 0)-- (2,-2);
\draw[color=blue, thick, ->](0, 0)-- (1,-2);
\draw[color=green, thick, ->](0, 0)-- (-2,2);
\draw[color=green, thick, ->](0, 0)-- (-2,3);
\filldraw[red] (1,2) circle (3pt)node[anchor=south]{$\rho$};
\node  at (2,0) [anchor=west]{$l_+$};
\node  at (0,2) [anchor=south]{$l_0'$};
\node  at (-2,-2) [anchor=east]{$l_-$};
\end{tikzpicture}\qquad
\begin{tikzpicture}[scale=0.4]
\draw[color=blue, thick,->](0, 0)-- (2, 0);
\draw[color=black, thick, ->] (0, 0)-- (0,2);
\draw[color=green, thick, ->](0, 0)-- (-2,-2);
\draw[color=blue, thick, ->](0, 0)-- (2,2);
\draw[color=blue, thick, ->](0, 0)-- (2,3);
\draw[color=blue, thick, ->](0, 0)-- (2,-2);
\draw[color=blue, thick, ->](0, 0)-- (1,-2);
\draw[color=green, thick, ->](0, 0)-- (-2,2);
\draw[color=green, thick, ->](0, 0)-- (-2,3);
\filldraw[red] (-1,2) circle (3pt)node[anchor=south]{$\rho'$};
\node  at (2,0) [anchor=west]{$l_+$};
\node  at (0,2) [anchor=south]{$l_0'$};
\node  at (-2,-2) [anchor=east]{$l_-$};
\end{tikzpicture}
\newline\newline
The change of the intersection number is due to the change of the cones which contribute to the sum.  In our case  we need to track  the contribution from the cones   Cone$(\vec{l}_0', \vec{l} )$ for all $\vec{l} \in S_\rho$. The intersection at $\rho$ includes cones  Cone$(\vec{l}_0', \vec{l}_+)$, depicted in blue on the picture,  while the intersection at $\rho'$ includes cones     Cone$(\vec{l}_0', \vec{l}_-)$ depicted in green.  The difference is 
\be
S_\rho \cdot S_0'  -S_{\rho'} \cdot S'_0 =  \sum_{ \vec{l}_+}    | \vec{l}_+ \times {\vec{l}_0}'| - \sum_{\vec{l}_-}    | \vec{l}_- \times {\vec{l}_0}'|\;\;.
\ee 
From the picture we see that all vectors $\vec{l}_+$ are related to $\vec{l}_0'$ by a counterclockwise rotation, while all $\vec{l}_-$ by a clockwise rotation, hence 
\be
 \vec{l}_+ \times {\vec{l}_0}' >0,\;\;\;   \vec{l}_- \times {\vec{l}_0}' <0\;\;.
\ee
We can rewrite the difference of intersection numbers 
\be
S_\rho \cdot S_0'  -S_{\rho'} \cdot S'_0 =  \sum_{ \vec{l}_+}     \vec{l}_+ \times {\vec{l}_0}'  -  \sum_{\vec{l}_-}    -( \vec{l}_- \times {\vec{l}_0}')  =   \sum_{\vec{l}\in S}  (\vec{l}\times \vec{l}_0') = 0.
\ee 
The last equality is due to the balancing condition (\ref{balanced_star}) for  the  star $S$.  $\hfill\blacksquare$
\newline\newline
We can use  relation (\ref{int_rays_cone}) to rewrite  sum  (\ref{int_stars_cone}) over indicator  functions on cones as  a sum over  intersection points of corresponding rays    to give an  enumerative description for the intersection number.  Hence an intersection number $S_\rho \cdot S_0'$ becomes  the weighted sum over intersection points  $p\in S_\rho\cap S_0'$ for pairs of  corresponding rays. The weight at intersection point $p$ is equal to the 
absolute value of the cross product for directional  vectors $\vec{l}_p$ and $\vec{l}'_p$ of the two rays intersecting at $p$, i.e. 
\be \label{star_intersection}
S_\rho \cdot S_0' =\sum_{p\; \in\; S_\rho\cap S_0'}  |\vec{l}_p\times \vec{l}_p'|.
\ee
{\bf Example}: On the picture below we present  the intersection of two stars. There are three intersection points and we zoomed in the circled  region around on of the  points and labeled the vectors of the two rays intersection at this point. 
 The absolute value of the wedge product for  the two rays  of the circled point equal  to one. the same true for the remaining two points. Hence we conclude that the intersection number for two stars equals to 3.
 \newline\newline
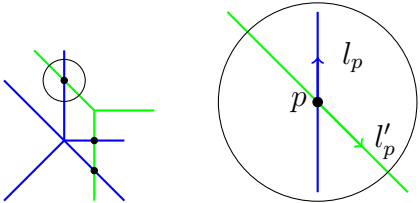
\begin{tikzpicture}[scale=0.4]
\draw[color=blue, thick](0, 0)-- (2, 0);
\draw[color=blue, thick] (0, 0)-- (0,3);
\draw[color=blue, thick](0, 0)-- (-2,-2);
\draw[color=blue, thick](0, 0)-- (2,-2);
\draw[color=blue, thick](0, 0)-- (-2,2);
\draw[color=green, thick](1, 1)-- (1,-2);
\draw[color=green, thick](1, 1)-- (3,1);
\draw[color=green, thick](1, 1)-- (-1,3);
\filldraw[black] (0,2) circle (3pt);
\filldraw[black] (1,0) circle (3pt);
\filldraw[black] (1,-1) circle (3pt);
\draw (0,2) circle (0.7);
\end{tikzpicture}\qquad
\begin{tikzpicture}[scale=0.6]
\draw[color=blue, thick](0, -2)-- (0, 2);
\draw[color=blue, thick, ->] (0, 0)-- (0,1);
\draw[color=green, thick](2, -2)-- (-2,2);
\draw[color=green, thick, ->](0, 0)-- (1,-1);
\filldraw[black] (0,0) circle (3pt)node[anchor=east]{$p$};
\filldraw (0.3,1) node[anchor=west]{$l_p$};
\filldraw (1.5,-1.5) node[anchor=south]{$l'_p$};
\draw (0,0) circle (2.2);
\end{tikzpicture}
\newline\newline
{\bf Remark}: The enumerative expression for the intersection of stars  can be naturally extended to the  intersection of two tropical curves.   For more details see Mikhalkin \cite{Mikhalkin2004,Mikh,mikhalkin2009tropical}.
\newline\newline
{\bf Remark}:  We can refine a self-intersection  for tropical curve $\vec{\G}$ from being defined only on cohomology classes to the representative. The  self-intersection  number for a curve $\vec{\G}$ is the weighted  union of vertex points  $V(\vec{\G})$.

\section{Tropical mirror for toric surfaces}

In this section we will adopt  the construction of the tropical  mirror  from \cite{Losev:2022tzr,Losev:2023bhj} to toric surfaces.  In particular,  we will describe the mirror superpotential, mirror states, holomorphic germs and tropical good section.

\subsection{Mirror relation}\label{sec_mirror_rel}

The mirror of the  complex   toric  surface $X$  is a non-compact $2$-dimensional Calabi-Yau  $X^{\vee} = \mathbb{C}^{\ast 2} $ with holomorphic  superpotential.   We will used  the  toric  representation $\mathbb{C}^{\ast 2} = \mathbb{R}^2 \times \mathbb{T}^2$ with radial coordinates $r^j$  and angular (holomorphic) coordinates $Y_j$.  The holomorphic top form in these coordinates 
\be\label{hol_top_form}
\Omega = dY_1\wedge dY_2.
\ee
  The mirror superpotential is
\be\label{mirr_superpotential}
W_X = \sum_{\vec{b}\in B_X}  q_{\vec{b}}\;\; e^{i \<\vec{b},\vec{Y}\>}.
\ee
where   we used the  pairing 
\be
\<\vec{b}, Y\> = b^1 Y_1+ b^2 Y_2.
\ee
The form (\ref{hol_top_form}) is invariant under $SL(2, \mathbb{Z})$, the linear transformations  with determinant equal to one and integer coefficients.   Let us arrange   vectors of the fan $B_X$ in  a counter-clockwise order and label them 
$\vec{b}_1, \vec{b}_2,...$.  A smooth projective toric variety   is a collection of smooth cones  Cone$(\vec{b}_k, \vec{b}_{k+1})$, i.e. cones   with $|\vec{b}_k \times \vec{b}_{k+1}|=1$.  Hence,  we can use an $SL(2, \mathbb{Z})$-rotation to rotate the pair of vectors $\vec{b}_1, \vec{b}_2$ to the standard basis of $\mathbb{Z}^2$, i.e.
\be
\vec{b}_1 \to (1,0),\;\;\; \vec{b}_{2} \to (0,1),\;\; \vec{b}_{k} \to \vec{b}'_{k},\;\; k>2.
\ee 
The superpotential in new basis becomes 
\be
W_X = q_{\vec{b}_1}\; e^{i Y_1} +  q_{\vec{b}_{2}} \;e^{i Y_2} + \sum_{k>2} q_{\vec{b}_k}\; e^{i \<\vec{b}_k', Y\>}.
\ee
The holomorphic top  form (\ref{hol_top_form}) is also invariant under constant shifts of $Y$-variables, hence we can use  
\be
Y_1 \to Y_1 -i \ln q_{\vec{b}_1},\;\;\; Y_2 \to Y_2 -i \ln q_{\vec{b}_{2}}
\ee
to simplify the  superpotential  into 
\be\label{plane_comp_superpotential}
W_X = e^{i Y_1} +  e^{i Y_2} + \sum_{k>2}  q'_{\vec{b}_k}\; e^{i \<\vec{b}_k', Y\>}.
\ee
The  new toric moduli   
\be
q'_{\vec{b}_k} = \frac{q_{\vec{b}_k}} {q^{b_k^1}_{\vec{b}_1} \cdot q^{b_k^2}_{\vec{b}_{2}}}
\ee
refine  Kahler moduli of $X$. If we formally set all Kahler moduli to zero we arrive into superpotential for the non-compact toric variety $\mathbb{C}^2$. 
Hence, the superpotential in the form (\ref{plane_comp_superpotential})  describes toric variety $X$  as a  compactification of $\mathbb{C}^2$.

\subsection{Mirror states and holomorphic germs}

{\bf Definition}: The Jacobi ring  for superpotential $W$ is
\be\label{def_Jacobi_ring}
J_W = R_{\mathbb{C}^{\ast 2}} / I_W,
\ee
where $R_{\mathbb{C}^{\ast 2}}$ is the ring  of holomorphic functions on $\mathbb{C}^{\ast 2}$. In our coordinates $R_{\mathbb{C}^{\ast 2}}$ is the ring of periodic functions of $Y$. The    $I_W$ is the  ideal generated by  the  partial derivatives  of  $W$
\be
I_W = \left\{  \frac{\p W}{\p Y_j} \right\}.
\ee
Let us consider a  graded  vector space of Landau-Ginzburg-Saito theory  
\be\label{def_lsg_vect_space}
V_{LGS} = R_{\mathbb{C}^{\ast 2}} \otimes \mathbb{C}[\psi_\Phi^i]
\ee
for parity-odd variables $\psi_\Phi^i$. On $V_{LGS}$ there is a  pair of graded-commuting differentials 
\be
 {\bf Q}_W = \frac{\p W}{\p Y_j} \frac{\p}{\p \psi_\Phi^j} ,\;\;\; {\bf G}_-  =\frac{\p}{\p Y_j} \frac{\p}{\p \psi_\Phi^j}.
\ee
The mirror map for observables is the map   from  the  de Rahm cohomology of toric space $X$   to  $({\bf Q}_W+z {\bf G_-})$-cohomology, i.e
\be
\Phi: H_{dR}^\ast(X)  \to H^\ast ({\bf Q}_W+z {\bf G_-}) : \gamma \mapsto \Phi_\g.
\ee
The mirror map is constructed in the following way: We turn an observable $\g$ into A-type  HTQM state $\Psi_\g$, then   construct the  corresponding mirror state $\Psi_\g^X$ and take its holomorphic germ $\Phi_\g$.

Let us   introduce  the   notation  $\Psi^{\vec{b}}$  for dressing of a state  $\Psi$ by a single compactifying divisor, labeled by   $\vec{b}$, i.e.
\be
\Psi^{\vec{b}} =  2\pi K G_- \mu_2 (\Psi_{\vec{b}}, \Psi) = 2\pi \int e^{-tH}dt\; G_+G_-\mu_2 (\Psi_{\vec{b}},  \Psi). 
\ee
The double dressing  by vectors $\vec{b}_1, \vec{b}_2$ in these notations  is 
\be
\Psi^{\vec{b}_1, \vec{b_2}} =  2\pi K G_- \mu_2 (\Psi_{\vec{b}_2}, \Psi^{\vec{b}_1}). 
\ee
The mirror state on the toric surface is given by 
\be
\Psi_\g^X = \Psi_\omega + \sum_{\vec{b}_1 \in B_X} \Psi^{\vec{b}_1}_\g + \sum_{\vec{b}_1, \vec{b}_2 \in B_X} \Psi^{\vec{b}_1, \vec{b}_2}_\g\;.
\ee
The holomorphic germ $ \Phi_\g$ for a mirror state $\Psi^X_\g$ is lowest component in $\psi$-expansion evaluated at $\vec{r}=0$, i.e
\be
 \Phi_\g = \Psi^X_\g \Big|_{\psi=0, r=0} .
\ee

\subsection{Tropical good section}\label{sec_good_section}

The construction of Jacobi ring comes with canonical projection $\pi_W: R_{\mathbb{C}^{\ast 2} }\to J_W$. Given a pair of homolorphic functions $\Phi_1$  and $\Phi_2 $ we can project their  product $\Phi_1 \Phi_2$ to the class $\pi_W(\Phi_1\Phi_2)$ in Jacobi ring $J_W$.  The section (which inverts $\pi_W$) $S_W: J_W \to R_{\mathbb{C}^{\ast 2}}$ turns this class into holomorphic function $S_W\; \pi_W(\Phi_1\Phi_2)$. The difference 
\be
\Phi_1\Phi_2 - S_W\; \pi_W(\Phi_1\Phi_2) 
\ee
 is trivial in Jacobi ring.  An isomorphism between the $J_W$  and $H^\ast ({\bf Q}_W)$ means that   there exists a map ${\bf \Sigma}_W: R_{\mathbb{C}^{\ast 2}} \to V_{LGS}$ such that 
\be
\Phi_1 \Phi_2 - S_W \pi_W (\Phi_1 \Phi_2) = {\bf Q}_W {\bf \Sigma}_W(\Phi_1\Phi_2),
\ee
and 
\be
{\bf \Sigma}_W S_W  =  0.
\ee
The choice of such ${\bf \Sigma}_W$ is known as the choice of homotopy for ${\bf Q}_W$.
\newline\newline
{\bf Definition}: We define a contact term  fo  $\Phi_1$ and $\Phi_2$  in LGS theory with  section $S_W$
\be\label{def_conn_good_section}
C^{S}_W (\Phi_1, \Phi_2) =  {\bf G}_- {\bf \Sigma}_W(\Phi_1\Phi_2).  
\ee
In other terms the product of two functions $\Phi_1\Phi_2$ can be decomposed into  the sum of the image of $S_W$ and  a linear combination of $\p^1 W,\p^2 W$, i.e.
\be\label{decom_s_w_sigma}
\Phi_1\Phi_2 = S_W \pi_W( \Phi_1\Phi_2)+\sigma_k \p^k W 
\ee
The ${\bf \Sigma}_W(\Phi_1\Phi_2)$ has the form $\sigma_k(Y) \psi_{\Phi}^k$, so  ${\bf G}_-$-action on it is 
\be
{\bf G}_-{\bf \Sigma}_W(\Phi_1\Phi_2) = \frac{\p\sigma_k(Y) }{\p Y_k},
\ee
i.e. just a divergence of the vector field $\sigma_k(Y) \p_{Y_k}$. Note that for a given $S_W$ the decomposition  in  (\ref{decom_s_w_sigma}) does not uniquely fixes the $\sigma_k(Y)$. The freedom of choice $\sigma$ is fixed by the choice of
 homotopy ${\bf \Sigma}_W$.

Note that the dependence of contact term $C_W$ on the choice of homotopy ${\bf \Sigma}_W$ is $({\bf Q}_W + z {\bf G}_-)$-exact. It was shown that the  correlation functions are well-defined in $H^\ast ({\bf Q}_W + z {\bf G}_-)$, so the choice of homotopy does not affect the recursion formula.

The tropical good for Landau-Ginzburg-Saito theory  is a  linear space  spanned by  identity germ $\Phi^X_1 =1$,  point germ $\Phi^X_\rho$, germs $\Phi^X_S$ for a basis in a space of stars.
\be
\hbox{Im} \; S^{trop} = \mathbb{C}\<1, \Phi^X_\rho, \Phi^X_{S}\> .
\ee

\subsection{Mirror state  for  point observable}

 The A-model state  for the $U(1)^2$-invariant Poincare-dual of  the point evaluation observable located at a  point   $\rho$ is   
\be
\Psi_{\rho} = \frac{1}{(2\pi)^2}\delta^2 (\vec{r}-\vec{\rho}\; )\; \psi_\Phi^1\psi_R^1  \psi_\Phi^2\psi_R^2.
\ee
 The single dressing of the state $\Psi_{\rho} $  by a divisor state is  
\be
\begin{split}
  \Psi_\rho^{\vec{b}}  & = 2\pi \int e^{-tH}dt\; G_+G_-\mu_2 (\Psi_{\vec{b}},  \Psi_\rho)  \\
&= \frac{1}{2\pi} q_{\vec{b}}\;e^{i \<\vec{b}, Y\>}    ( \vec{b}\times \vec{\psi}_R) ( \vec{b}\times \vec{\psi}_\Phi) \int^\infty_0  dt\;  \delta^2 (\vec{r} -\vec{\rho}- \vec{b}t ) .  
\end{split}
\ee
We used  
\be
G_- \left(  \psi_{\Phi}^2 \psi_{\Phi}^1 \;e^{i \<\vec{b}, Y\>}\right)   =  (b^2 \psi^1_{\Phi} - b^1 \psi^2_\Phi) \;e^{i \<\vec{b}, Y\>} = (\vec{\psi}_\Phi \times \vec{b})\;e^{i \<\vec{b}, Y\>}
\ee
and similar relation for  $G_+$. The integral of a delta function implies that  the  single dressed state  $  \Psi_\rho^{\vec{b}}$ has support on the ray $R_{b,\rho}$. Moreover,  the inclusion of  $\psi$-dependence describes the  $\Psi_\rho^{\vec{b}}$ as the multiple of  the state for  Poincare-dual (\ref{Poincare_dual_ray}) of the ray $R_{b,\rho}$,  hence we can write
\be
\Psi_\rho^{\vec{b}} = q_{\vec{b}}\;e^{i \<\vec{b}, Y\>}  \Psi_{R_{b,\rho}}\;\;.
\ee
We can represent   the dressing of the state  $\Psi_\rho$  by all divisors from the fan $B_X$, i.e. 
\be
\sum_{\vec{b}\in B_X} \Psi_\rho^{\vec{b}} =\sum_{\vec{b}\in B_X}  q_{\vec{b}}\;e^{i \<\vec{b}, Y\>}  \Psi_{R_{b,\rho}}
\ee
as the evaluation  state for the quasi-star (no balancing condition) $S_\rho$  with rays identical to the rays of  $B_X$, equipped with   holomorphic functions. The  ray along vector $\vec{b}$ is equipped with the function    $q_{\vec{b}}\;e^{i \<\vec{b}, Y\>}$.

The dressing of $\Psi_{\rho}$  by two divisor states 
\be
\begin{split}
\Psi^{\vec{b}_1, \vec{b_2}}_\rho & = q_{\vec{b}_1}q_{\vec{b}_2} \;e^{i \<\vec{b}_1+\vec{b}_2, Y\>}  (\vec{b}_1\times \vec{b}_2)^2 \int^\infty_0 \int^\infty_0  dt_1 dt_2\;  \delta (\vec{r}-\vec{\rho}-\vec{b}_1 t_1-(\vec{b}_1+\vec{b}_2)t_2 )  \\
& = q_{\vec{b}_1}q_{\vec{b}_2} \; e^{i \<\vec{b}_1+\vec{b}_2, Y\>}  |\vec{b}_1\times \vec{b}_2| \chi_{ \vec{b}_1,  \vec{b}_1+\vec{b}_2}(\vec{r} -\vec{\rho}\;).  
\end{split}
\ee
We used an integral representation (\ref{int_rep_theta_cone})  for indicator function on a cone  and 
\be
 \vec{b}_1\times( \vec{b}_2 +\vec{b}_1) =  \vec{b}_1\times \vec{b}_2.
\ee
Note that  the dressing is not symmetric under exchange of  $\vec{b}_1$ and $\vec{b}_2$, because the  indicator functions  have support at different regions, i.e 
\be
\chi_{\vec{b}_1, \vec{b}_1+\vec{b}_2} (\vec{r}\;) \neq \chi_{ \vec{b}_2, \vec{b}_1+\vec{b}_2}(\vec{r}\;).
\ee
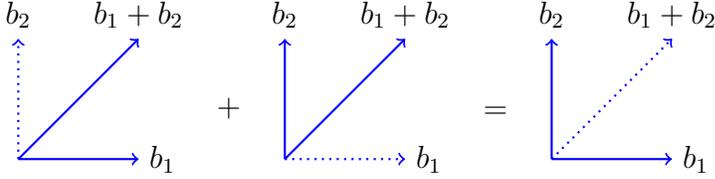
\begin{tikzpicture}[scale=0.4]
\draw[color=blue, thick,->](0, 0)-- (4, 0);
\draw[color=blue, thick, dotted, ->] (0, 0)-- (0,4);
\draw[color=blue, thick, ->](0, 0)-- (4,4);
\node  at (4,0) [anchor=west]{$b_1$};
\node  at (0,4) [anchor=south]{$b_2$};
\node  at (4,4) [anchor=south]{$b_1+b_2$};
\node  at (7,1) [anchor=south]{$+$};
\end{tikzpicture}
\begin{tikzpicture}[scale=0.4]
\draw[color=blue, thick, dotted , ->](0, 0)-- (4, 0);
\draw[color=blue, thick,  ->] (0, 0)-- (0,4);
\draw[color=blue, thick, ->](0, 0)-- (4,4);
\node  at (4,0) [anchor=west]{$b_1$};
\node  at (0,4) [anchor=south]{$b_2$};
\node  at (4,4) [anchor=south]{$b_1+b_2$};
\node  at (7,1) [anchor=south]{$=$};
\end{tikzpicture}
\begin{tikzpicture}[scale=0.4]
\draw[color=blue, thick,->](0, 0)-- (4, 0);
\draw[color=blue, thick, ->] (0, 0)-- (0,4);
\draw[color=blue, thick, dotted,->](0, 0)-- (4,4);
\node  at (4,0) [anchor=west]{$b_1$};
\node  at (0,4) [anchor=south]{$b_2$};
\node  at (4,4) [anchor=south]{$b_1+b_2$};
\end{tikzpicture}
\newline
 We can notice that two  orders of   performing the  double dressing   have the same holomorphic function, so we can naturally simplify the sum using an equality 
\be
\chi_{\vec{b}_1, \vec{b}_1+\vec{b}_2}(\vec{r}\;)+  \chi_{ \vec{b}_2, \vec{b}_1+\vec{b}_2}(\vec{r}\;) = \chi_{ \vec{b}_1, \vec{b}_2}(\vec{r}\;)\;\;\;.
\ee 
The graphical reprsenation of this equality is given  on the picture above.
The  holomorphic germ  for the point observable    is
\be\label{pt_mirr_hol_germ}
\Phi_\rho^X   =\Psi^{\vec{b}_1, \vec{b_2}}_\rho\Big|_{ r=0} =   \frac12  \sum_{\vec{b}, \vec{b}' \in B_X} |\vec{b}\times \vec{b}'| \; q_{\vec{b}}q_{\vec{b}'}\;e^{\<\vec{b}+\vec{b}', Y\>}  \chi_{ \vec{b}, \vec{b}'}(-\vec{\rho}\;)\;.
\ee 
Our construction for the holomorphic germ gives different  holomorphic  functions depending on the location of the point $\rho$ on $\mathbb{R}^2$.
However,  diffent  holomorphic germs  represent the same  cohomology class.
\newline\newline
{\bf Proposition (cone crossing)}:  Holomorphic germs   (\ref{pt_mirr_hol_germ})  represent the same class in $ ({\bf Q}_W + z {\bf G}_-)$-cohomology for all values of $\rho$.
\newline\newline
{\bf Proof}: The  holomorphic germ   (\ref{pt_mirr_hol_germ}) changes each time  the point $\rho$ crosses cone of the fan  $B_X$. Let us consider the change of  the germ  under crossing of a single vector $\vec{b}_0$. On the picture below we colored in  blue all vectors $\vec{b}_+$ such that the    cones Cone$(\vec{b}_0, \vec{b}_+)$ give non-zero contribution to the $\Phi_{-\rho}^X$.    We colored in green all vectors $\vec{b}_-$, such that the    cones Cone$(\vec{b}_0, \vec{b}_-)$ contribute to $\Phi_{-\rho'}^X$.
\newline\newline
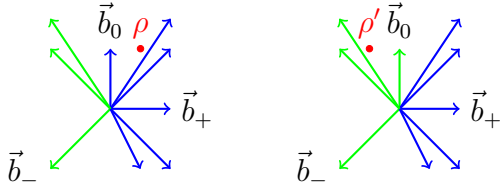
\begin{tikzpicture}[scale=0.4]
\draw[color=blue, thick,->](0, 0)-- (2, 0);
\draw[color=blue, thick, ->] (0, 0)-- (0,2);
\draw[color=green, thick, ->](0, 0)-- (-2,-2);
\draw[color=blue, thick, ->](0, 0)-- (2,2);
\draw[color=blue, thick, ->](0, 0)-- (2,3);
\draw[color=blue, thick, ->](0, 0)-- (2,-2);
\draw[color=blue, thick, ->](0, 0)-- (1,-2);
\draw[color=green, thick, ->](0, 0)-- (-2,2);
\draw[color=green, thick, ->](0, 0)-- (-2,3);
\filldraw[red] (1,2) circle (3pt)node[anchor=south]{$\rho$};
\node  at (2,0) [anchor=west]{$\vec{b}_+$};
\node  at (0,2) [anchor=south]{$\vec{b}_0$};
\node  at (-2,-2) [anchor=east]{$\vec{b}_-$};
\end{tikzpicture}\qquad
\begin{tikzpicture}[scale=0.4]
\draw[color=blue, thick,->](0, 0)-- (2, 0);
\draw[color=green, thick, ->] (0, 0)-- (0,2);
\draw[color=green, thick, ->](0, 0)-- (-2,-2);
\draw[color=blue, thick, ->](0, 0)-- (2,2);
\draw[color=blue, thick, ->](0, 0)-- (2,3);
\draw[color=blue, thick, ->](0, 0)-- (2,-2);
\draw[color=blue, thick, ->](0, 0)-- (1,-2);
\draw[color=green, thick, ->](0, 0)-- (-2,2);
\draw[color=green, thick, ->](0, 0)-- (-2,3);
\filldraw[red] (-1,2) circle (3pt)node[anchor=south]{$\rho'$};
\node  at (2,0) [anchor=west]{$\vec{b}_+$};
\node  at (0,2) [anchor=south]{$\vec{b}_0$};
\node  at (-2,-2) [anchor=east]{$\vec{b}_-$};
\end{tikzpicture}
\newline\newline
The  difference between two functions  is given by 
\be
\Phi_{-\rho}^X -\Phi_{-\rho'}^X = \sum_{\vec{b}_+}   | \vec{b}_+\times \vec{b}_0| \;q_{\vec{b}_+}q_{\vec{b}_0}\;e^{\<\vec{b}_0+\vec{b}_+, Y\>} -\sum_{\vec{b}_-}  |\vec{b}_-\times \vec{b}_0| \;q_{\vec{b}_0}q_{\vec{b}_-}\;e^{i\<\vec{b}_0+\vec{b}_-, Y\>}.
\ee
All vectors $\vec{b}_+$ are related to $\vec{b}_0$ by a counterclockwise rotation, while $\vec{b}_-$ are related by clockwise rotation hence
\be
\vec{b}_+\times \vec{b}_0 >0,\;\;\; \vec{b}_-\times \vec{b}_0<0,
\ee
 and we can rewrite 
\be
\begin{split}
\Phi_{-\rho}^X -\Phi_{-\rho'}^X  &=  \sum_{\vec{b}_+}   ( \vec{b}_+\times \vec{b}_0) \;q_{\vec{b}_+}q_{\vec{b}_0}\;e^{\<\vec{b}_0+\vec{b}_+, Y\>} -\sum_{\vec{b}_-} -(\vec{b}_-\times \vec{b}_0)\; q_{\vec{b}_0}q_{\vec{b}_-}\;e^{i\<\vec{b}_0+\vec{b}_-, Y\>}\\
 & = \sum_{\vec{b} \in B_X}  (\vec{b}\times \vec{b}_0) \;q_{\vec{b}}q_{\vec{b}_0}\;e^{\<\vec{b}+\vec{b}_0, Y\>} .
 \end{split}
\ee
We can   express the sum  ver $\vec{b}$ as a derivative of superpotential   (\ref{mirr_superpotential}), i.e.
\be
\begin{split}
\Phi_{-\rho}^X -\Phi_{-\rho'}^X   & = \sum_{\vec{b} \in B_X}  (\vec{b}\times \vec{b}_0) \;q_{\vec{b}}q_{\vec{b}_0}\;e^{\<\vec{b}+\vec{b}_0, Y\>}  =  q_{\vec{b}_0} \; e^{i \<\vec{b}_0, Y\>} (-i \overrightarrow{\p_Y W}\times \vec{b}_0 )  =  {\bf Q}_W\chi_{\vec{b}_0} ,
 \end{split}
\ee
for the  state 
\be
\chi_{\vec{b}_0} = -i q_{\vec{b}_0} \; e^{i\<\vec{b}_0, Y\>}  (\vec{\psi}_\Phi\times \vec{b}_0).
\ee
The state  $\chi_{\vec{b}_0}$  is  ${\bf G}_-$-closed, i.e.
\be
{\bf G}_-\chi_{\vec{b}_0} =\frac{\p}{\p Y_k} \frac{\p}{\p \psi_\Phi^k}\left(q_{\vec{b}_0} \; e^{i \<\vec{b}_0, Y\>}  (\vec{\psi}_\Phi\times \vec{b}_0) \right) =  iq_{\vec{b}_0} \; e^{i \<\vec{b}_0, Y\>}  (\vec{b}_0\times \vec{b}_0) = 0.
\ee
Hence the change of the holomorphic germ under the crossing of a single ray along $\vec{b}_0$ is  exact, i.e
\be
\Phi_{-\rho}^X -\Phi_{-\rho'}^X = ({\bf Q}_W + z {\bf G}_-)\chi_{\vec{b}_0} .
\ee
For general translation from $\rho$  to   $\rho'$ we may  need to perform several cone-crossings.  $\hfill\blacksquare$
\newline\newline
{\bf Proposition (Enumerative expression for germs)}: The holomorphic germ for the point observable can be constructed in the following way: We  perform the weighted sum over the intersection points $p$ of the fan   $B_X$ at origin and the fan  $-B_X^{-\rho}$ at point $-\rho$. The weights are determined by the direction vectors of the intersecting rays.  
\be\label{enumer_hol_germ_point}
\Phi^X_\rho  =\frac12  \sum_{p\; \in\; B_X\cap -B_X^{-\rho}} q_{\vec{b}_p}q_{\vec{b}_p'} |\vec{b}_p\times \vec{b}_p'| \;e^{i \<\vec{b}_p + \vec{b}'_p, Y\>}.
\ee
\newline\newline
\begin{tikzpicture}[scale=0.4]
\draw[color=blue, thick,->](0, 0)-- (4, 0);
\draw[color=blue, thick, ->] (0, 0)-- (0,4);
\draw[color=blue, thick, ->](0, 0)-- (-2,-2);
\draw[color=blue, thick, ->](0, 0)-- (4,4);
\node  at (-4,2) [anchor=west]{$B_X$};
\end{tikzpicture}\qquad\qquad
\begin{tikzpicture}[scale=0.4]
\draw[color=green, thick,->](0, 0)-- (-4, 0);
\draw[color=green, thick, ->] (0, 0)-- (0,-4);
\draw[color=green, thick, ->](0, 0)-- (-4,-4);
\draw[color=green, thick, ->](0, 0)-- (2,2);
\node  at (-4,2) [anchor=west]{$-B_X$};
\end{tikzpicture}\qquad\qquad
\begin{tikzpicture}[scale=0.4]
\draw[color=blue, thick,->](0, 0)-- (4, 0);
\draw[color=blue, thick, ->] (0, 0)-- (0,4);
\draw[color=blue, thick, ->](0, 0)-- (-2,-2);
\draw[color=blue, thick, ->](0, 0)-- (4,4);
\draw[color=green, thick,->](0+3, 0+2)-- (-4+3, 0+2);
\draw[color=green, thick, ->] (0+3, 0+2)-- (0+3,-4+2);
\draw[color=green, thick, ->](0+3, 0+2)-- (-4+3,-4+2);
\draw[color=green, thick, ->](0+3, 0+2)-- (2+3,2+2);
\filldraw[black] (2,2) circle (3pt);
\filldraw[black] (1,0) circle (3pt);
\filldraw[black] (0,2) circle (3pt);
\filldraw[black] (3,0) circle (3pt);
\node  at (0,0) [anchor=east]{$0$};
\node  at (0+3,0+2) [anchor=west]{$-\rho$};
\end{tikzpicture}
\newline\newline
{\bf Proof}: The relation  (\ref{ray_intersect})  tells us   that the  step functions  in a sum  (\ref{pt_mirr_hol_germ})  describe the    intersection points  two rays: one from $B_X$, while the other from $B_X^{-\rho}$. 
Hence the sum over  step functions  is the same as the sum over intersection points. The multiplicative factors in front of the  indicator  functions   give us the weight factors in (\ref{enumer_hol_germ_point}). Hence the proof is complete. $\hfill\blacksquare$

\subsection{Mirror state  for   star-observable} \label{sec_star_mirr}

The  A-model  state $\Psi_{R_{l,\rho}}$  for a ray $R_{l, \rho}$ is  constructed from the the  Poincare-dual  form (\ref{Poincare_dual_ray})  by replacement $dr \to \psi_R$ and $d\phi \to \psi_\Phi$. Namely,
\be
\Psi_{R_{l,\rho}} = \frac{1}{2\pi}(\vec{l}\times \vec{\psi}_R)(\vec{l}\times \vec{\psi}_\Phi)  \int^\infty_0  dt \; \delta^2(\vec{r}-\vec{\rho}- t \;\vec{l}). 
\ee
The single divisor dressing of the state 
\be
\Psi_{R_{l,\rho}}^{\vec{b}} =  (\vec{l}\times \vec{b})^2q_{\vec{b}}\; e^{i \<\vec{b}, Y\>}\int^\infty_0 dt \int^\infty_0 ds\;\; \delta^{2} (\vec{r}-\vec{\rho}-s\; \vec{l}- t\; \vec{b}) .
\ee
The integral is the  step  function on a cone (\ref{ray_intersect}), hence we can further simplify 
\be
\Psi_{R_{l, \rho}}^{\vec{b}} =  |\vec{l}\times \vec{b}| \; q_{\vec{b}}\;e^{i \<\vec{b}, Y\>} \; \chi_{\vec{l}, \vec{b}}(\vec{r}-\vec{\rho}\;).
\ee
The mirror state for the ray-observable 
\be
\Psi_{R_{l, \rho}} ^{X} =\Psi_{R_{l, \rho}} +  \sum_{\vec{b}\in B_X} \Psi_{R_{l, \rho}}^{\vec{b}}  =\Psi_{R_{l, \rho}}  + \sum_{\vec{b}\in B_X}    |\vec{l}\times \vec{b}|\; q_{\vec{b}}\; e^{i \<\vec{b}, Y\>} \chi_{\vec{l}, \vec{b}}(\vec{r}-\vec{\rho}\;).
\ee
The A-model state for the star observable $S_\rho$ is a sum of the states for its rays, i.e. 
\be
\Psi_{S_\rho} = \sum_{\vec{l} \in S}  \Psi_{R_{l, \rho}}, 
\ee
while corresponding mirror state 
\be
\Psi_{S_\rho}^{X} =  \Psi_{S_\rho} + \sum_{\vec{l} \in S}  \sum_{\vec{b} \in B_X}   |\vec{l}\times \vec{b}|\; q_{\vec{b}}\; e^{i \<\vec{b}, Y\>} \chi_{\vec{l}, \vec{b}}(\vec{r}-\vec{\rho}\;).
\ee
The holomorphic germ for the star observable $S_\rho$ is 
\be\label{star_hol_germ}
\Phi_{S_\rho}^{X}  = \Psi_{S_\rho}^{X}\Big|_{\psi=r=0}  = \sum_{\vec{l} \in S}  \sum_{\vec{b} \in B_X}    |\vec{l}\times \vec{b}| \;q_{\vec{b}}\; e^{i \<\vec{b}, Y\>} \chi_{\vec{l}, \vec{b}}(-\vec{\rho}\;).
\ee
{\bf Proposition}:  Holomorphic germs $\Phi_{S_\rho}^{X}$ in  (\ref{star_hol_germ}) represent the same class in $ ({\bf Q}_W + z {\bf G}_-)$-cohomology for all values of $\rho$.
\newline\newline
{\bf Proof}: The holomorphic germ (\ref{star_hol_germ}) changes each time  the point $\rho$ crosses either   ray of    fan $B_X$  or ray of star $S_\rho$. Let us consider the change of function under crossing the vector $\vec{l}_0\in S_\rho$. On the picture below we colored in  blue all vectors $\vec{b}_+$, such that the    cones Cone$(\vec{l}_0, \vec{b}_+)$ give non-zero contribution to  $\Phi_{S_{-\rho}}^X$.    We colored in green all vectors $\vec{b}_-$, such that the    cones Cone$(\vec{l}_0, \vec{b}_-)$ contribute to $\Phi_{S_{-\rho'}}^X$.
\newline\newline
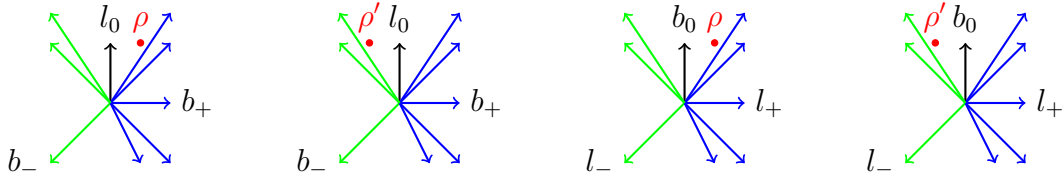
\begin{tikzpicture}[scale=0.4]
\draw[color=blue, thick,->](0, 0)-- (2, 0);
\draw[color=black, thick, ->] (0, 0)-- (0,2);
\draw[color=green, thick, ->](0, 0)-- (-2,-2);
\draw[color=blue, thick, ->](0, 0)-- (2,2);
\draw[color=blue, thick, ->](0, 0)-- (2,3);
\draw[color=blue, thick, ->](0, 0)-- (2,-2);
\draw[color=blue, thick, ->](0, 0)-- (1,-2);
\draw[color=green, thick, ->](0, 0)-- (-2,2);
\draw[color=green, thick, ->](0, 0)-- (-2,3);
\filldraw[red] (1,2) circle (3pt)node[anchor=south]{$\rho$};
\node  at (2,0) [anchor=west]{$b_+$};
\node  at (0,2) [anchor=south]{$l_0$};
\node  at (-2,-2) [anchor=east]{$b_-$};
\end{tikzpicture}\qquad
\begin{tikzpicture}[scale=0.4]
\draw[color=blue, thick,->](0, 0)-- (2, 0);
\draw[color=black, thick, ->] (0, 0)-- (0,2);
\draw[color=green, thick, ->](0, 0)-- (-2,-2);
\draw[color=blue, thick, ->](0, 0)-- (2,2);
\draw[color=blue, thick, ->](0, 0)-- (2,3);
\draw[color=blue, thick, ->](0, 0)-- (2,-2);
\draw[color=blue, thick, ->](0, 0)-- (1,-2);
\draw[color=green, thick, ->](0, 0)-- (-2,2);
\draw[color=green, thick, ->](0, 0)-- (-2,3);
\filldraw[red] (-1,2) circle (3pt)node[anchor=south]{$\rho'$};
\node  at (2,0) [anchor=west]{$b_+$};
\node  at (0,2) [anchor=south]{$l_0$};
\node  at (-2,-2) [anchor=east]{$b_-$};
\end{tikzpicture}\qquad
\begin{tikzpicture}[scale=0.4]
\draw[color=blue, thick,->](0, 0)-- (2, 0);
\draw[color=black, thick, ->] (0, 0)-- (0,2);
\draw[color=green, thick, ->](0, 0)-- (-2,-2);
\draw[color=blue, thick, ->](0, 0)-- (2,2);
\draw[color=blue, thick, ->](0, 0)-- (2,3);
\draw[color=blue, thick, ->](0, 0)-- (2,-2);
\draw[color=blue, thick, ->](0, 0)-- (1,-2);
\draw[color=green, thick, ->](0, 0)-- (-2,2);
\draw[color=green, thick, ->](0, 0)-- (-2,3);
\filldraw[red] (1,2) circle (3pt)node[anchor=south]{$\rho$};
\node  at (2,0) [anchor=west]{$l_+$};
\node  at (0,2) [anchor=south]{$b_0$};
\node  at (-2,-2) [anchor=east]{$l_-$};
\end{tikzpicture}\qquad
\begin{tikzpicture}[scale=0.4]
\draw[color=blue, thick,->](0, 0)-- (2, 0);
\draw[color=black, thick, ->] (0, 0)-- (0,2);
\draw[color=green, thick, ->](0, 0)-- (-2,-2);
\draw[color=blue, thick, ->](0, 0)-- (2,2);
\draw[color=blue, thick, ->](0, 0)-- (2,3);
\draw[color=blue, thick, ->](0, 0)-- (2,-2);
\draw[color=blue, thick, ->](0, 0)-- (1,-2);
\draw[color=green, thick, ->](0, 0)-- (-2,2);
\draw[color=green, thick, ->](0, 0)-- (-2,3);
\filldraw[red] (-1,2) circle (3pt)node[anchor=south]{$\rho'$};
\node  at (2,0) [anchor=west]{$l_+$};
\node  at (0,2) [anchor=south]{$b_0$};
\node  at (-2,-2) [anchor=east]{$l_-$};
\end{tikzpicture}
\newline\newline
Hence, the  difference between two  germs  is given by 
\be
\Phi_{S_{-\rho}}^X -\Phi_{S_{-\rho'}}^X=  \sum_{\vec{b}_+}    |\vec{b}_+\times \vec{l}_0|\; q_{\vec{b}_+}\; e^{i \<\vec{b}_+, Y\>} - \sum_{ \vec{b}_-}    |\vec{b}_-\times \vec{l}_0| \; q_{\vec{b}_-}\; e^{i \<\vec{b}_-, Y\>}.
\ee
All vectors  $\vec{b}_+$  are related to $\vec{l}_0$  by a counterclockwise rotation, while $\vec{b}_-$ by a clockwise one, hence 
\be
\vec{b}_+\times \vec{l}_0>0,\;\; \vec{b}_-\times \vec{l}_0<0.
\ee 
We can simplify the absolute values in the sum 
\be
\begin{split}
\Phi_{S_{-\rho}}^X -\Phi_{S_{-\rho'}}^X & =  \sum_{\vec{b}_+}    (\vec{b}_+\times \vec{l}_0)\; q_{\vec{b}_+}\; e^{i \<\vec{b}_+, Y\>} - \sum_{ \vec{b}_-}  -  (\vec{b}_-\times \vec{l}_0) \; q_{\vec{b}_-}\; e^{i \<\vec{b}_-, Y\>}\\
 & =  \sum_{\vec{b} \in B_X}  (\vec{b}\times \vec{l}_0)\; q_{\vec{b}} \;e^{\<\vec{b}, Y\>}.
\end{split}
\ee
We can   express the sum  ver $\vec{b}$ as a derivative of superpotential   (\ref{mirr_superpotential}), i.e.
\be
\begin{split}
  \sum_{\vec{b} \in B_X}  (\vec{b}\times \vec{l}_0)\; q_{\vec{b}} \;e^{\<\vec{b}, Y\>} & =  -i \overrightarrow{\p_Y W_X}\times \vec{l}_0  ={\bf Q}_W\chi_{\vec{l}_0},
 \end{split}
\ee
for the state 
\be
\chi_{\vec{l}_0} =-i( \vec{\psi}_\Phi\times \vec{l}_0).
\ee
The state $\chi_{\vec{l}_0} $  is ${\bf G}_-$-closed, i.e.
\be
{\bf G}_-\chi_{\vec{l}_0} = \frac{\p}{\p Y_k} \frac{\p}{\p \psi_\Phi^k}\left( \vec{\psi}_\Phi\times \vec{l}_0 \right)  = 0,
\ee
hence 
\be
  \Phi_{S_{-\rho}}^X -\Phi_{S_{-\rho'}}^X  = ({\bf Q}_W +z{\bf G}_-)\chi_{\vec{l}_0}.
\ee
The other possibility is  the crossing of  some vector $\vec{b}_0 \in B_X$. The  difference between two functions  is given by Cone$(\vec{l}, \vec{b}_0)$-contributions. We can repeat the analysis for orientations to simplify the absolute values 

\be
\begin{split}
\Phi_{S_{-\rho}}^X -\Phi_{S_{-\rho'}}^X &=  \sum_{ \vec{l}_+}    |\vec{l}_+\times \vec{b}_0| \;q_{\vec{b}_0}\; e^{i \<\vec{b}_0, Y\>} - \sum_{\vec{l}_-}    |\vec{l}_-\times \vec{b}_0|\; q_{\vec{b}_0}\; e^{i \<\vec{b}_0, Y\>} \\
&=  \sum_{ \vec{l}_+}    (\vec{l}_+\times \vec{b}_0 )\;q_{\vec{b}_0}\; e^{i \<\vec{b}_0, Y\>} - \sum_{\vec{l}_-}   - (\vec{l}_-\times \vec{b}_0)\; q_{\vec{b}_0}\; e^{i \<\vec{b}_0, Y\>} \\
& =     \;q_{\vec{b}_0}\; e^{i \<\vec{b}_0, Y\>}  \sum_{\vec{l}\in S}  (\vec{l}\times \vec{b}_0) = 0.
\end{split}
\ee 
The last equality is due to the balancing condition (\ref{balanced_star}) for the  star-observable.  

The general translation of  the star $S$ from $\rho$ to $\rho'$ can be split into finitely many   crossings of a single vector either from $S$ of $B_X$. Since single crossing preserves the class of holomorphic germ in  $({\bf Q}_W + z {\bf G}_-)$-cohomology, so it is true  for the finitely-many crossings. $\hfill\blacksquare$

\subsection{Mirror for tropical curve observable}

We can use a relation (\ref{int_rays_cone})   to replace  indicator functions  by intersection points of rays  to  give an  enumerative  formulation for the holomorphic germ (\ref{star_hol_germ}) for   a star observable:
The  sum over the intersection points  of a star $S_\rho$  and the reflection of the fan $-B_X$, weighted  cross-product of corresponding vectors and  holomorphic function $q_{\vec{b}}\; e^{i \<\vec{b}, Y\>}$.   
\be
\Phi^X_{S_\rho} = \sum_{p \;\in\; S_{\rho}\cap -B_X}   |\vec{l}_p\times \vec{b}_p|\; q_{\vec{b}}\; e^{i \<\vec{b}_p, Y\>}.
\ee
\newline\newline
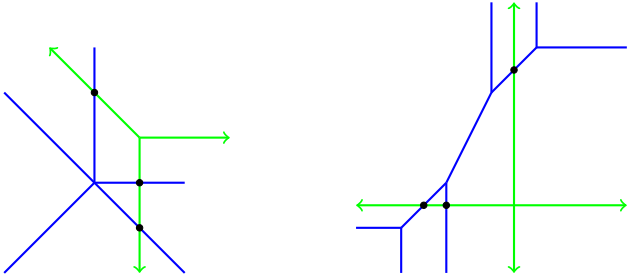
\begin{tikzpicture}[scale=0.6]
\draw[color=blue, thick](0, 0)-- (2, 0);
\draw[color=blue, thick] (0, 0)-- (0,3);
\draw[color=blue, thick](0, 0)-- (-2,-2);
\draw[color=blue, thick](0, 0)-- (2,-2);
\draw[color=blue, thick](0, 0)-- (-2,2);
\draw[color=green, thick, ->](1, 1)-- (1,-2);
\draw[color=green, thick, ->](1, 1)-- (3,1);
\draw[color=green, thick, ->](1, 1)-- (-1,3);
\filldraw[black] (0,2) circle (2pt);
\filldraw[black] (1,0) circle (2pt);
\filldraw[black] (1,-1) circle (2pt);
\end{tikzpicture}\qquad\qquad
\begin{tikzpicture}[scale=0.6]
\draw[color=blue, thick] (-2, -1) -- (-1, -1);
\draw[color=blue, thick]  (-1, -1)-- (-1,-2);
\draw[color=blue, thick](-1, -1)-- (0,0);
\draw[color=blue, thick](0,0)--(0,-2);
\draw[color=blue, thick](0,0)-- (1,2)--(1,4);
\draw[color=blue, thick](1,2)--(2,3)--(2,4);
\draw[color=blue, thick](2,3) -- (4,3);
\draw[color=green, thick, <->](1.5, -2)-- (1.5,4);
\draw[color=green, thick, <->](-2, -0.5)-- (4,-0.5);
\filldraw[black] (1.5,2.5) circle (2pt);
\filldraw[black] (0,-.5) circle (2pt);
\filldraw[black] (-0.5,-0.5) circle (2pt);
\end{tikzpicture} 
\newline\newline
We can generalize the formula for the holomorphic germ  from  the star (maximally degenerate tropical curve)  to an arbitrary tropical curve (possibly of higher genus).   Namely, 
\be\label{curve_mirr_germ}
\Phi^X_\G = \sum_{p \;\in\; \G\cap - B_X}   |\vec{l}_p\times \vec{b}_p|\; q_{\vec{b}_p}\; e^{i \<\vec{b}_p, Y\>}.
\ee
Each intersection point  $p$ is an intersection of a ray along $\vec{b}_p$ from $B_X$  and an edge of a graph, representing tropical curve $\G$ equipped  with integer vector $\vec{l}_p$. 
\newline\newline
{\bf Proposition}:  The class of  holomorphic germ $\Phi^X_\G $ in $({\bf Q}_W + z {\bf G}_-)$-cohomology  is independent of the moduli of tropical curve.
\newline\newline
{\bf Proof}: There are two types of events which are  can change the holomorphic germ  as we change the moduli of the tropical curve: 
\begin{itemize}
\item {\bf Ray of $B_X$ crosses  vertex  of $\G$}:  The change in holomorphic  germ   is controlled  by the cones Cone$(\vec{b}, \vec{l})$ for vector $\vec{b}$ on the ray of $B_X$  and  vectors $\vec{l}$  connecting to   the vertex $V$ of $\G$. The analysis of the change is identical to the analysis we did for stars in section \ref{sec_star_mirr}.  In particular,  the change was proportional to the balancing condition for a star.  In the case of tropical curve the difference will be proportional to the balancing condition for  the vertex $V$, i.e.
\be
\Phi_\G^X - \Phi_{\G'}^X =  q_{\vec{b}} \; e^{i \<\vec{b}, Y\>}\;   \sum_{\vec{l} \in V} \left(  \vec{l} \times  \vec{b} \right)=0.
\ee
The last equality is due to  balancing condition  which holds for all vertices   of tropical curve $\G$. For more details see Mikhalkin \cite{Mikhalkin2004,Mikh,mikhalkin2009tropical}

\item {\bf Edge of $\G$ crosses vertex of $B_X$}:  The change in holomorphic  germ   is controlled  by the cones Cone$(\vec{b}, \vec{l})$ for vector $\vec{l}$ assigned to   the edge of $\G$  and  vectors $\vec{b}$  from the fan of $X$. The analysis of the change is identical to the analysis we   did for  stars in section \ref{sec_star_mirr}.  In particular,  the change of a holomorphic germ is given by 
\be
\Phi_\G^X - \Phi_{\G'}^X = ({\bf Q}_W + z {\bf G}_-)\chi_{\vec{l}}\;,\;\;\; \chi_{\vec{l}} = -i( \vec{\psi}_\Phi  \times \vec{l}). 
\ee
\end{itemize} 
Both events change the holomorphic germ by at most  $({\bf Q}_W + z {\bf G}_-)$-exact term, hence preserve the cohomology class of a germ.   Any change of moduli for a tropical curve is a chain of the finitely many crossing events, hence the proof is complete.     $\hfill\blacksquare$
\newline\newline
{\bf Example}: On the pictures below we present the intersection of the green  tropical curve   and the toric  fan  depicted in blue.  Each consecutive picture describe a translation of the toric fan to the left. First, second  and  third picture describe a crossing for the (vertical) ray of  the fan   and vertices of the curve.  The holomorphic  germ does not change since the intersection points lie on the very same rays of the green curve and the cross of corresponding ray-vectors are the same. On the fourth  and fifth pictures we observe the crossing of the green vertex and blue edges of the tropical curve. The intersection  points on horizontal rays of the tropical fan 
move from one ray to another. Hence the holomorphic germ changes. 
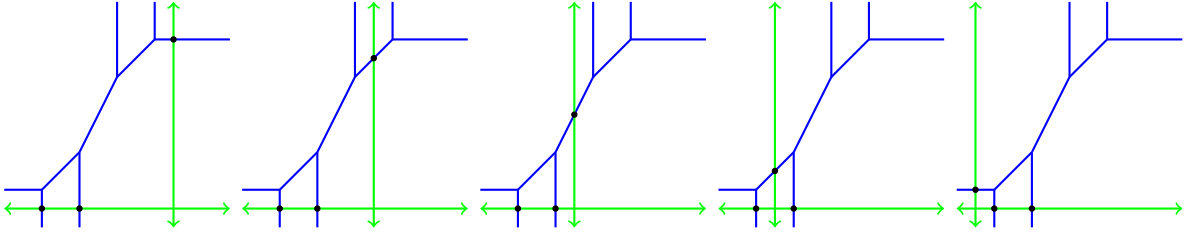
\newline\newline
\begin{tikzpicture}[scale=0.5]
\draw[color=blue, thick] (-2, -1) -- (-1, -1);
\draw[color=blue, thick]  (-1, -1)-- (-1,-2);
\draw[color=blue, thick](-1, -1)-- (0,0);
\draw[color=blue, thick](0,0)--(0,-2);
\draw[color=blue, thick](0,0)-- (1,2)--(1,4);
\draw[color=blue, thick](1,2)--(2,3)--(2,4);
\draw[color=blue, thick](2,3) -- (4,3);
\draw[color=green, thick, <->](2.5, -2)-- (2.5,4);
\draw[color=green, thick, <->](-2, -1.5)-- (4,-1.5);
\filldraw[black] (2.5,3) circle (2pt);
\filldraw[black] (0,-1.5) circle (2pt);
\filldraw[black] (-1,-1.5) circle (2pt);
\end{tikzpicture} 
\begin{tikzpicture}[scale=0.5]
\draw[color=blue, thick] (-2, -1) -- (-1, -1);
\draw[color=blue, thick]  (-1, -1)-- (-1,-2);
\draw[color=blue, thick](-1, -1)-- (0,0);
\draw[color=blue, thick](0,0)--(0,-2);
\draw[color=blue, thick](0,0)-- (1,2)--(1,4);
\draw[color=blue, thick](1,2)--(2,3)--(2,4);
\draw[color=blue, thick](2,3) -- (4,3);
\draw[color=green, thick, <->](1.5, -2)-- (1.5,4);
\draw[color=green, thick, <->](-2, -1.5)-- (4,-1.5);
\filldraw[black] (1.5,2.5) circle (2pt);
\filldraw[black] (0,-1.5) circle (2pt);
\filldraw[black] (-1,-1.5) circle (2pt);
\end{tikzpicture} 
\begin{tikzpicture}[scale=0.5]
\draw[color=blue, thick] (-2, -1) -- (-1, -1);
\draw[color=blue, thick]  (-1, -1)-- (-1,-2);
\draw[color=blue, thick](-1, -1)-- (0,0);
\draw[color=blue, thick](0,0)--(0,-2);
\draw[color=blue, thick](0,0)-- (1,2)--(1,4);
\draw[color=blue, thick](1,2)--(2,3)--(2,4);
\draw[color=blue, thick](2,3) -- (4,3);
\draw[color=green, thick, <->](0.5, -2)-- (0.5,4);
\draw[color=green, thick, <->](-2, -1.5)-- (4,-1.5);
\filldraw[black] (0.5,1) circle (2pt);
\filldraw[black] (0,-1.5) circle (2pt);
\filldraw[black] (-1,-1.5) circle (2pt);
\end{tikzpicture} 
\begin{tikzpicture}[scale=0.5]
\draw[color=blue, thick] (-2, -1) -- (-1, -1);
\draw[color=blue, thick]  (-1, -1)-- (-1,-2);
\draw[color=blue, thick](-1, -1)-- (0,0);
\draw[color=blue, thick](0,0)--(0,-2);
\draw[color=blue, thick](0,0)-- (1,2)--(1,4);
\draw[color=blue, thick](1,2)--(2,3)--(2,4);
\draw[color=blue, thick](2,3) -- (4,3);
\draw[color=green, thick, <->](-0.5, -2)-- (-0.5,4);
\draw[color=green, thick, <->](-2, -1.5)-- (4,-1.5);
\filldraw[black] (-0.5,-0.5) circle (2pt);
\filldraw[black] (0,-1.5) circle (2pt);
\filldraw[black] (-1,-1.5) circle (2pt);
\end{tikzpicture} 
\begin{tikzpicture}[scale=0.5]
\draw[color=blue, thick] (-2, -1) -- (-1, -1);
\draw[color=blue, thick]  (-1, -1)-- (-1,-2);
\draw[color=blue, thick](-1, -1)-- (0,0);
\draw[color=blue, thick](0,0)--(0,-2);
\draw[color=blue, thick](0,0)-- (1,2)--(1,4);
\draw[color=blue, thick](1,2)--(2,3)--(2,4);
\draw[color=blue, thick](2,3) -- (4,3);
\draw[color=green, thick, <->](-1.5, -2)-- (-1.5,4);
\draw[color=green, thick, <->](-2, -1.5)-- (4,-1.5);
\filldraw[black] (-1.5,-1) circle (2pt);
\filldraw[black] (0,-1.5) circle (2pt);
\filldraw[black] (-1,-1.5) circle (2pt);
\end{tikzpicture}

\section{Divisor relation }

\subsection{Divisor relation  for Gromov-Witten invariants}
Let us recall the following property of the Gromov-Witten invariants. For a hypersurface $H$  with Poincare-dual form  $\g_H$ and classes  $\g_1,..,\g_n \in H^\ast (X)$ the following relation holds 
\be\label{divisor_rel}
\begin{split}
\< \g_H, \g_1, ..,\g_n\>_{0,\beta}^X  =\left( \int_{\Sigma_\beta} \g_H\right) \cdot  \<\g_1, ..,\g_n\>_{0,\beta}^X,
\end{split}
\ee 
where $\beta $ is the degree of curve and $\Sigma_{\beta}$  is a curve representing class $\beta$   in the  Kahler cone of $H_2(X)$. 

Let us give an equivalent formulation of the divisor relation  for tropical mirror of toric surfaces.  The hyperplane $H$ in 2 dimensions   becomes a  tropical curve. Moreover,  we can turn a  tropical curve into a 
star by shrinking the lengths of internal edges. A  tropical curve and the corresponding star are in the same cohomology class on $X$, hence without loss of generality we will assume that $H$ is a star.
Our expression (\ref{star_intersection})  for the intersection number implies that   two stars $S, S'$  have  positive intersection  number $S \cdot S'$. Stars form a cone under the union operation: we can take a union of two stars.  Equivalently,  we can add the  corresponding Poincare-dual forms
\be
\g_{S\cup S'} = \g_S+ \g_{S'}.
\ee 
Hence, we can use  stars to  define a Kahler cone for $X$ and express the intersection number in (\ref{divisor_rel}) as an intersection number for  two  stars
\be
\int_{\Sigma_\beta} \g_H = S_\beta \cdot H. 
\ee
The weighted sum over the Gromov-Witten invariants    can be  written 
\be
\sum_\beta  \< \g_1, ..,\g_n\>_{0,\beta}^X \; q^\beta  =\sum_{S_\beta}  \<\g_1, ..,\g_n\>_{0,\beta}^X \; \prod_{\vec{l} \in S_\beta} q_{\vec{l}}\;\;\;,
\ee 
where $q_{\vec{l}}$  are toric moduli associated to the rays $\vec{l}$ of a star $S_\beta$.

\subsection{Tropical divisor relation from  LGS}

 The tropical  mirror \cite{Losev:2023bhj}  allows us to express the weighted sum in terms of correlation function in B-type HTQM
\be
\sum_\beta  q^\beta \< \g_S, \g_1, ..,\g_n\>_{0,\beta}^X  =  \< \Psi^X_{\g_S},  \Psi^X_{\g_1}, ..,\Psi_{\g_n}^X\>_{Q_W}.
\ee
We can replace the mirror state for a star observable  $\Psi^X_{\g_S}$  by its holomorphic germ $\Phi_{S}$  and use the recursion  relation in B-type HTQM to arrive into  
\be\label{div_recursion_relation}
  \< \Psi^{X}_{\g_S},  \Psi^{X}_{\g_1}, ..,\Psi_{\g_n}^{X}\>_{Q_W} = \<\Phi_{S},  \Psi^X_{\g_1}, ..,\Psi_{\g_n}^X\>_{Q_{W^X}} =   \frac{d}{d\e}\Big|_{\e=0}\<  \Psi^\e_{\g_1}, ..,\Psi^\e_{\g_n}\>_{Q_{W^X_\e}},
\ee
for deformed superpotential 
\be
W^\e_X = W_X + \e \;\Phi^X_S,
\ee
and deformed states 
\be
 \Psi^\e_{\g_k} =  \Psi^{W_X^\e}_{\g_k},\;\;\; k=1,...,n.
\ee
The holomorphic germ of the mirror state for the star observable  
\be
 \Phi^X_{S_\rho}  = \sum_{ \vec{l} \in S}   \sum_{\vec{b}\in B_X}    |\vec{l} \times \vec{b}| \; q_{\vec{b}}\; e^{i \<\vec{b}, Y\>}  \chi_{\vec{l}, \vec{b}}(-\rho)
\ee
gives us the  deformed superpotential  
\be
W^\e_X (q_{\vec{b}}) = \sum_{\vec{b}\in B_X} q_{\vec{b}} \; e^{i \< \vec{b}, Y\>}  + \e\sum_{ \vec{l} \in S}   \sum_{\vec{b}\in B_X}    |\vec{l} \times \vec{b}| \; q_{\vec{b}}\; e^{i \<\vec{b}, Y\>}  \chi_{\vec{l}, \vec{b}}(-\rho)=W_X (q_{\vec{b}}^\e).
\ee
The deformed superpotential has  the same $e^{i \<\vec{m}, Y\>}$-terms, but with $\e$-dependent coefficients.   Hence, it  describes the same toric fan $B_X$, but modified toric moduli. In particular,  moduli before and after deformation are related by the multiplicative factor 
\be
\begin{split}
q_{\vec{b}}^\e  &= q_{\vec{b}} \left(1+ \e  \sum_{ \vec{l} \in S}    |\vec{l} \times \vec{b}| \; \chi_{\vec{l}, \vec{b}}(-\rho) \right).
\end{split}
\ee
The recursion relation (\ref{div_recursion_relation}) is an equality between polynomials in $q$,   which implies equality between coefficients  of corresponding monomials. Let us look at coefficients for monomial
\be
q^\beta = \prod_{\vec{b} \in S_\beta} q_{\vec{b}}.
\ee 
Note that the monomials   $(q^\e)^\beta$ and $q^\beta$ are related by the multiplicative factor
\be
(q^\e)^\beta=\prod_{\vec{b} \in S_\beta}  q_{\vec{b}}^\e = q^\beta \prod_{\vec{b}\in S_\beta}\left(1+ \e  \sum_{ \vec{l} \in S}    |\vec{l} \times \vec{b}| \; \chi_{\vec{l}, \vec{b}}(-\rho) \right).
\ee
Hence the coefficients  in polynomial expansion are related by the multiplicative factor as well.

The coefficient in the expansion of the first expression in (\ref{div_recursion_relation}), by construction,    is the tropical $(n+1)$-pont Gromov-Witten invariant,  while the coefficients of the last expression 
in (\ref{div_recursion_relation})   is the multiple of $n$-point  Gromov-Witten invariant. In particular 
\be
  \<\g_S, \g_1, ..,\g_n\>_{0,\beta}^X =   \<\g_1, ..,\g_n\>_{0,\beta}^X \;  \frac{d}{d\e}\Big|_{\e=0} \prod_{\vec{b}\in S_\beta}  \left(1+ \e  \sum_{ \vec{l} \in S}    |\vec{l} \times \vec{b}| \; \chi_{\vec{l}, \vec{b}}(-\rho) \right).
\ee
The derivative evaluates into the intersection number for stars
\be
\sum_{ \vec{b}\in S_\beta,\;\; \vec{l} \in S}    |\vec{l} \times \vec{b}| \; \chi_{\vec{l}, \vec{b}}(-\rho) = S_\beta \cdot S.
\ee
Hence, we derived the divisor relation for tropical Gromov-Witten invariants on toric surface from the recursion formula in B-type HTQM.

\section{Mirror for selected toric surfaces}

\subsection{ $\mathbb{P}^2$}

 The compactification   polytope  for   $\mathbb{P}^2$ and the corresponding fan are presented below  
 \newline\newline
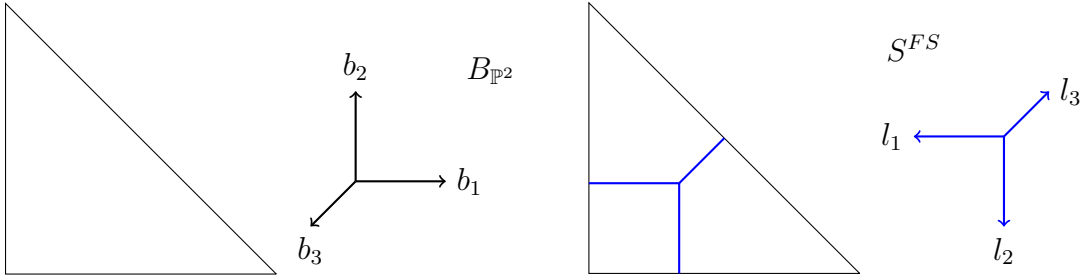
\begin{tikzpicture}[scale=0.6]
\draw(4,-2)-- (-2, -2) -- (-2, 4)--(4,-2);
\end{tikzpicture}
\begin{tikzpicture}[scale=0.6]
\draw[color=black, thick,->](0, 0)-- (2, 0);
\draw[color=black, thick, ->] (0, 0)-- (0,2);
\draw[color=black, thick, ->](0, 0)-- (-1,-1);
\node  at (2,0) [anchor=west]{$b_1$};
\node  at (0,2) [anchor=south]{$b_2$};
\node  at (-1,-1) [anchor=north]{$b_3$};
\node  at (3,3) [anchor=north]{$B_{\mathbb{P}^2}$};
\end{tikzpicture}\qquad
\begin{tikzpicture}[scale=0.6]
\draw(4,-2)-- (-2, -2) -- (-2, 4)--(4,-2);
\draw[color=blue, thick] (-2, 0) -- (0, 0)-- (0,-2);
\draw[color=blue, thick](0, 0)-- (1,1);
\end{tikzpicture}
\begin{tikzpicture}[scale=0.6]
\draw[color=blue, thick,->](0, 0)-- (-2, 0);
\draw[color=blue, thick, ->] (0, 0)-- (0,-2);
\draw[color=blue, thick, ->](0, 0)-- (1,1);
\node  at (-2,0) [anchor=east]{$l_1$};
\node  at (0,-2) [anchor=north]{$l_2$};
\node  at (1,1) [anchor=west]{$l_3$};
\node  at (-2,2.5) [anchor=north]{$S^{FS}$};
\end{tikzpicture}
\newline\newline
The  fan for  $\mathbb{P}^2$ is generated by three vectors 
\be
B_{\mathbb{P}^2} = \{ \vec{b}_1= (1,0), \vec{b}_2=(0,1),\vec{b}_3= (-1,-1)\}.
\ee
 Since  $\dim H_{dR}^2(\mathbb{P}^2) =1$ we have a single hyperplane observable. The natural choice is to use the Fubini-Study  form on $\mathbb{P}^2$.  The tropical limit of the Fubini-Study form  is the star observable 
\be
S^{FS} = \{ \vec{l}_1 = - \vec{b}_1,  \vec{l}_2 =  -\vec{b}_2,  \vec{l}_3 =  -\vec{b}_3\}.
\ee
The mirror superpotential  (\ref{mirr_superpotential}) is
\be
 W_{\mathbb{P}^2}  =  q_1\; e^{i Y_1} + q_2\; e^{i Y_2}  + q_3\; e^{-i Y_1-iY_2}. 
\ee
The mirror state  is 
\be\label{FS_mirror_p2}
\begin{split}
 \Psi_{S^{FS}}^{\mathbb{P}^2}  &=  \Psi_{S^{FS}} + \sum_{\vec{b} \in B_X}   \sum_{\vec{l} \in S^{FS}}    |\vec{l}\times \vec{b}|\; q_{\vec{b}}\; e^{i \<\vec{b}, Y\>} \chi_{\vec{l}, \vec{b}} (\vec{r}-\vec{\rho}) \\
 & =\Psi_{S^{FS}} +  q_1\;e^{i  \<\vec{b}_1,Y\>} (    \chi_{\vec{l}_2, \vec{b}_1}  + \chi_{\vec{l}_3, \vec{b}_1} )  + q_2\; e^{i  \<\vec{b}_2,Y\>} (    \chi_{\vec{l}_1, \vec{b}_2} +    \chi_{\vec{l}_3, \vec{b}_2} )+ q_3\; e^{i  \<\vec{b}_3,Y\>} (    \chi_{\vec{l}_1, \vec{b}_3} +  \chi_{\vec{l}_2, \vec{b}_3} )  \\
 & =\Psi_{S^{FS}} +  q_1\;e^{i  \<\vec{b}_1,Y\>}     \chi_{\vec{l}_2, \vec{l}_3} (\vec{r}-\vec{\rho})     + q_2\; e^{i  \<\vec{b}_2,Y\>}   \chi_{\vec{l}_1, \vec{l}_3} (\vec{r}-\vec{\rho})  + q_3\; e^{i  \<\vec{b}_3,Y\>}     \chi_{\vec{l}_1, \vec{l}_2}  (\vec{r}-\vec{\rho}), 
 \end{split}
\ee
where we used  cross  product between non-parallel pairs 
\be
|\vec{l} \times \vec{b}| =1.
\ee
There are three    holomorphic germs of the mirror state (\ref{FS_mirror_p2}), labeled by three cones 
\be
\Phi_{S^{FS}_{-\rho} }  = \left\{
\begin{array}{ll}
  q_3\; e^{i \<\vec{b}_3, Y\>} = q_3\; e^{-i Y_1-iY_2}, & \vec{\rho} \in \hbox{Cone}(\vec{l}_1, \vec{l}_2);  \\
 q_1\; e^{i \<\vec{b}_1, Y\>} = q_1\; e^{i Y_1}, & \vec{\rho} \in \hbox{Cone}(\vec{l}_2, \vec{l}_3); \\
 q_2\; e^{i \<\vec{b}_2, Y\>} = q_2\; e^{i Y_2}, & \vec{\rho} \in \hbox{Cone}(\vec{l}_1, \vec{l}_3).
 \end{array}
\right.
\ee
 The enumerative description  of the holomorphic germ for the  star observable $S$ is constructed from the diagrams below 
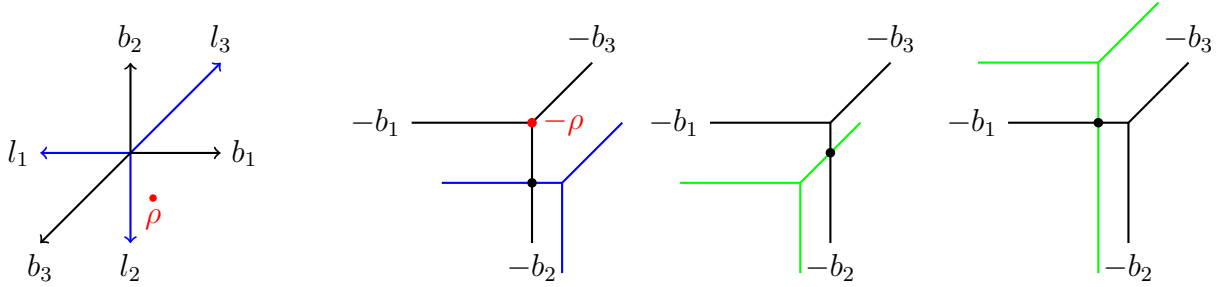
\newline\newline
\begin{tikzpicture}[scale=0.6]
\draw[color=blue, thick,->](0, 0)-- (-2, 0);
\draw[color=blue, thick, ->] (0, 0)-- (0,-2);
\draw[color=blue, thick, ->](0, 0)-- (2,2);
\node  at (-2,0) [anchor=east]{$l_1$};
\node  at (0,-2) [anchor=north]{$l_2$};
\node  at (2,2) [anchor=south]{$l_3$};
\draw[color=black, thick,->](0, 0)-- (2, 0);
\draw[color=black, thick, ->] (0, 0)-- (0,2);
\draw[color=black, thick, ->](0, 0)-- (-2,-2);
\node  at (2,0) [anchor=west]{$b_1$};
\node  at (0,2) [anchor=south]{$b_2$};
\node  at (-2,-2) [anchor=north]{$b_3$};
\filldraw[red] (0.5 ,-1) circle (2pt)node[anchor=north]{$\rho$};
\end{tikzpicture}\qquad\
\begin{tikzpicture}[scale=0.8]
\draw[color=black, thick] (-2, 0) -- (0, 0)-- (0,-2);
\draw[color=black, thick](0, 0)-- (1,1);
\node  at (-2,0) [anchor=east]{$-b_1$};
\node  at (0,-2) [anchor=north]{$-b_2$};
\node  at (1,1) [anchor=south]{$-b_3$};
\draw[color=blue, thick](0.5, -1)-- (1.5,0);
\draw[color=blue, thick](0.5, -1)-- (0.5,-2.5);
\draw[color=blue, thick](0.5, -1)-- (-1.5,-1);
\filldraw[black] (0,-1) circle (2pt);
\filldraw[red] (0 ,0) circle (2pt)node[anchor=west]{$-\rho$};
\end{tikzpicture}
\begin{tikzpicture}[scale=0.8]
\draw[color=black, thick] (-2, 0) -- (0, 0)-- (0,-2);
\draw[color=black, thick](0, 0)-- (1,1);
\node  at (-2,0) [anchor=east]{$-b_1$};
\node  at (0,-2) [anchor=north]{$-b_2$};
\node  at (1,1) [anchor=south]{$-b_3$};
\draw[color=green, thick](-0.5, -1)-- (0.5,0);
\draw[color=green, thick](-0.5, -1)-- (-0.5,-2.5);
\draw[color=green, thick](-0.5, -1)-- (-2.5,-1);
\filldraw[black] (0,-0.5) circle (2pt);
\end{tikzpicture}
\begin{tikzpicture}[scale=0.8]
\draw[color=black, thick] (-2, 0) -- (0, 0)-- (0,-2);
\draw[color=black, thick](0, 0)-- (1,1);
\node  at (-2,0) [anchor=east]{$-b_1$};
\node  at (0,-2) [anchor=north]{$-b_2$};
\node  at (1,1) [anchor=south]{$-b_3$};
\draw[color=green, thick](-0.5, 1)-- (0.5,2);
\draw[color=green, thick](-0.5, 1)-- (-0.5,-2.5);
\draw[color=green, thick](-0.5, 1)-- (-2.5,1);
\filldraw[black] (-0.5,0) circle (2pt);
\end{tikzpicture}
\newline\newline
All three  functions represent the same cohomology class i.e 
\be
\Phi_{S^{FS}_\rho} =  q_1\; e^{iY_1} =  q_2 \;e^{iY_2}=q_3\; e^{-iY_1-iY_2} \in H^\ast({\bf Q}_{\mathbb{P}^2}+z {\bf G}_-).
\ee
Indeed, we can perform the cone  crossings to determine the exact terms
\be
  \begin{split}
({\bf Q}_{\mathbb{P}^2}+z {\bf G}_-)(-i\psi^1_\Phi) &= -i\p_1  W  =q_1\;  e^{iY_1} - q_3\; e^{-iY_1 - iY_2}, \\
({\bf Q}_{\mathbb{P}^2}+z {\bf G}_-)(-i \psi^2_\Phi ) &=-i\p_2  W  =   q_2\; e^{iY_2} -q_3\;  e^{-iY_1-iY_2}.
  \end{split}
  \ee
The  three possible choices of holomorphic germs give us the following deformations of toric moduli 
\be
\begin{split}
\hbox{Cone}(\vec{l}_1, \vec{l}_2):\; (q_1, q_2, q_3) \to  (q_1, q_2, q_3 (1+\e) ),  \\
 \hbox{Cone}(\vec{l}_2, \vec{l}_3):\; (q_1, q_2, q_3) \to  (q_1(1+\e), q_2, q_3  ),  \\ 
 \hbox{Cone}(\vec{l}_1, \vec{l}_3):\; (q_1, q_2, q_3) \to  (q_1, q_2(1+\e), q_3 ).
\end{split}
\ee   The three deformations above describe the same Kahler moduli deformation. Indeed, the weight factor for the degree-d curves, written in terms of toric moduli  using the star representative
\be
q^\beta = (q_1 q_2 q_3)^d  =  (q_1 q_2 q_3)^d (1+\e)^d = (q_1 q_2 q_3)^d (1+d\cdot \e +\cO(\e^2)).
\ee
The $\cO(\e)$ terms in the last equality match with  intersection of degree star $S_\beta = d\;S^{FS}$  with a star $S = \e\; S^{FS}$. Indeed, we can evaluate  
\be
S_\beta  \cdot S = d \cdot\e \; S^{FS} \cdot S^{FS} = d\cdot \e.
\ee
In the last equality we used  the   self-intersection number for the Fubini-Study  star 
\be
S^{FS} \cdot S^{FS} = \sum_{\vec{l},\; \vec{l}' \in S^{FS}} |\vec{l}\times \vec{l}'| \;\chi_{\vec{l}, -\vec{l}'}(\rho) =1.
\ee
 There are three possible  germs for the mirror state of the point observable at point $-\vec{\rho}$,  labeled by three cones
 \be
\Phi_{-\rho }  = \left\{
\begin{array}{ll}
|\vec{b}_1\times \vec{b}_2|\; q_1 q_2\; e^{i \<\vec{b}_1+\vec{b}_2, Y\>} = q_1q_2\; e^{i Y_1+iY_2} , & \vec{\rho} \in \hbox{Cone}(\vec{b}_1, \vec{b}_2);  \\
 |\vec{b}_1\times \vec{b}_3|\; q_1 q_3\; e^{i \<\vec{b}_1+\vec{b}_3, Y\>}  = q_1q_3\; e^{-i Y_2} , & \vec{\rho} \in \hbox{Cone}(\vec{b}_2, \vec{b}_3); \\
 |\vec{b}_2\times \vec{b}_3|\; q_2 q_3\; e^{i \<\vec{b}_2+\vec{b}_3, Y\>}  = q_2q_3\; e^{-i Y_1} , & \vec{\rho} \in \hbox{Cone}(\vec{b}_1, \vec{b}_3).
 \end{array}
\right.
\ee
We can  perform the cone crossing  to derive the   relations 
\be
  \begin{split}
({\bf Q}_{\mathbb{P}^2}+z {\bf G}_-)(-i q_2 e^{iY_2}\psi^1_\Phi) &=-i  q_2 e^{iY_2} \p_1 W = q_1q_2 e^{iY_1+iY_2} - q_2q_3 e^{-iY_1 }, \\
({\bf Q}_{\mathbb{P}^2}+z {\bf G}_-) (-i q_1 e^{iY_1}\psi^2_\Phi ) &=-i q_1 e^{i Y_1} \p_2  W  =   q_1q_2 e^{iY_1+iY_2} -q_1 q_3 e^{-iY_2}.
  \end{split}
\ee
which  imply  that all three holomorphic germs  are in the same class 
\be
q_2 q_3\; e^{-iY_1} = q_1 q_3 \;e^{-iY_2} = q_1q_2\; e^{iY_1+iY_2} \in  H^\ast ({\bf Q}_{\mathbb{P}^2}+z {\bf G}_-).
\ee
Using holomorphic germs for trivial,  point and  hyperplane  observables we can  describe  the tropical good section
\be\label{trop_good_sec_p2}
\hbox{Im}\; S^{trop}_{\mathbb{P}^2} = \mathbb{C}\<1, \Phi_{S^{FS}}, \Phi_\rho\> = \mathbb{C}\<1, q_3 \;e^{-iY_1-i Y_2},q_1q_2\; e^{iY_1+i Y_2}\>.
\ee

\subsection{ $\mathbb{P}^1 \times \mathbb{P}^1$}

The compactifying polyhedron and the fan   for $\mathbb{P}^1 \times \mathbb{P}^1$ are  presented below
 \newline\newline
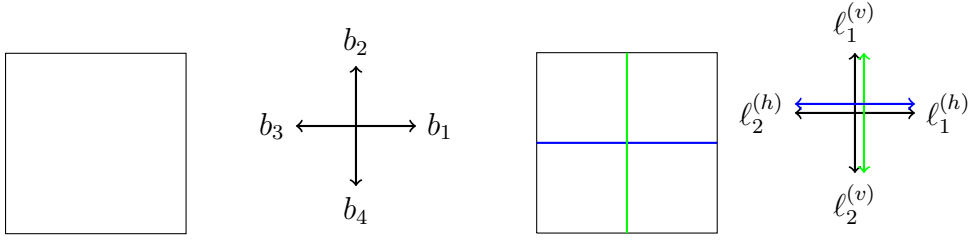
\begin{tikzpicture}[scale=0.6]
\draw(-2,-2)-- (2, -2) -- (2, 2)--(-2,2)--(-2,-2);
\end{tikzpicture}\qquad
\begin{tikzpicture}[scale=0.4]
\draw[color=black, thick,->](0, 0)-- (2, 0);
\draw[color=black, thick, ->] (0, 0)-- (0,2);
\draw[color=black, thick, ->](0, 0)-- (-2,0);
\draw[color=black, thick, ->](0, 0)-- (0,-2);
\node  at (2,0) [anchor=west]{$b_1$};
\node  at (0,2) [anchor=south]{$b_2$};
\node  at (-2,0) [anchor=east]{$b_3$};
\node  at (0,-2) [anchor=north]{$b_4$};
\end{tikzpicture}
\qquad
\begin{tikzpicture}[scale=0.6]
\draw(-2,-2)-- (2, -2) -- (2, 2)--(-2,2)--(-2,-2);
\draw[color=blue, thick] (-2, 0) -- (2, 0);
\draw[color=green, thick](0, -2)-- (0,2);
\end{tikzpicture}
\begin{tikzpicture}[scale=0.4]
\draw[color=black, thick,->](0, 0)-- (2, 0);
\draw[color=black, thick, ->] (0, 0)-- (0,2);
\draw[color=black, thick, ->](0, 0)-- (-2,0);
\draw[color=black, thick, ->](0, 0)-- (0,-2);
\draw[color=blue, thick,->](0, 0.3)-- (-2, 0.3);
\draw[color=green, thick, ->] (0.3, 0)-- (0.3,2);
\draw[color=blue, thick, ->](0, 0.3)-- (2,0.3);
\draw[color=green, thick, ->](0.3, 0)-- (0.3,-2);
\node  at (2,0) [anchor=west]{$\ell_1^{(h)}$};
\node  at (0,2) [anchor=south]{$\ell_1^{(v)}$};
\node  at (-2,0) [anchor=east]{$\ell_2^{(h)}$};
\node  at (0,-2) [anchor=north]{$\ell_2^{(v)}$};
\end{tikzpicture}
\newline\newline
The   generators   of the fan 
\be
B_{\mathbb{P}^1 \times \mathbb{P}^1} = \{ \vec{b}_1= (1,0), \vec{b}_2=(0,1),\vec{b}_3= (-1,0), \vec{b}_4= (0,-1)\}
\ee
give us the  mirror superpotential (\ref{mirr_superpotential}) of the form 
\be
W^{\mathbb{P}^1 \times \mathbb{P}^1} =  q_1\; e^{i Y_1} + q_2\; e^{i Y_2}+ q_3\; e^{-i Y_1} + q_4\; e^{-i Y_2}.
\ee
The $H^{2}_{dR} (\mathbb{P}^1 \times \mathbb{P}^1)$ is 2-dimensional  and we can use  the   Fubini-Study forms on $\mathbb{P}^1$-factors as a basis. The  tropical limit of the Fubini-Study forms is the pair of 2-ray stars: horizontal labeled by $h$, depicted in blue  and  vertical, labeled by $v$, depicted in  green  on the picture above. The corresponding A-model states
\be
\begin{split}
\Psi_{v}& = \delta (r^1) \; \psi_\Phi^1\psi_R^1, \\
\Psi_{h} &=  \delta (r^2)\;   \psi_\Phi^2\psi_R^2.
\end{split}
\ee
The  holomorphic germs  are determined from   the four intersections    depicted  below
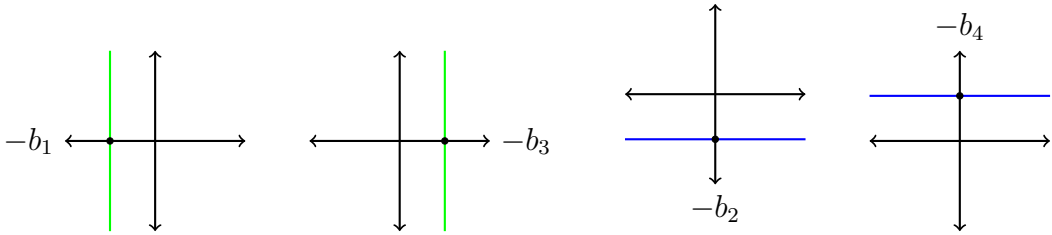
\newline\newline
\begin{tikzpicture}[scale=0.6]
\draw[color=black, thick,->](0, 0)-- (2, 0);
\draw[color=black, thick, ->] (0, 0)-- (0,2);
\draw[color=black, thick, ->](0, 0)-- (-2,0);
\draw[color=black, thick, ->](0, 0)-- (0,-2);
\draw[color=green, thick](-1, -2)-- (-1, 2);
\filldraw[black] (-1,0) circle (2pt);
\node  at (-2,0) [anchor=east]{$-b_1$};
\end{tikzpicture}\qquad
\begin{tikzpicture}[scale=0.6]
\draw[color=black, thick,->](0, 0)-- (2, 0);
\draw[color=black, thick, ->] (0, 0)-- (0,2);
\draw[color=black, thick, ->](0, 0)-- (-2,0);
\draw[color=black, thick, ->](0, 0)-- (0,-2);
\draw[color=green, thick](1, -2)-- (1, 2);
\filldraw[black] (1,0) circle (2pt);
\node  at (2,0) [anchor=west]{$-b_3$};
\end{tikzpicture}\qquad
\begin{tikzpicture}[scale=0.6]
\draw[color=black, thick,->](0, 0)-- (2, 0);
\draw[color=black, thick, ->] (0, 0)-- (0,2);
\draw[color=black, thick, ->](0, 0)-- (-2,0);
\draw[color=black, thick, ->](0, 0)-- (0,-2);
\draw[color=blue, thick](-2, -1)-- (2, -1);
\filldraw[black] (0,-1) circle (2pt);
\node  at (0,-2) [anchor=north]{$-b_2$};
\end{tikzpicture}\qquad
\begin{tikzpicture}[scale=0.6]
\draw[color=black, thick,->](0, 0)-- (2, 0);
\draw[color=black, thick, ->] (0, 0)-- (0,2);
\draw[color=black, thick, ->](0, 0)-- (-2,0);
\draw[color=black, thick, ->](0, 0)-- (0,-2);
\draw[color=blue, thick](-2, 1)-- (2, 1);
\filldraw[black] (0,1) circle (2pt);
\node  at (0,2) [anchor=south]{$-b_4$};
\end{tikzpicture}
\newline\newline
There is a single  intersection point in all four cases,  so  corresponding holomorphic contain single term. The straightforward evaluation gives us 
 \be
\Phi_{v} = \left\{
\begin{array}{ll}
 q_1\; e^{i \<\vec{b}_1, Y\>} = q_1\; e^{i Y_1} , & \rho^1<0;  \\
q_3\; e^{i \<\vec{b}_3, Y\>} = q_3\; e^{-i Y_1}, & \rho^1>0; \\
 \end{array}
\right.
\ee
and
 \be
\Phi_{h} = \left\{
\begin{array}{ll}
q_2\; e^{i \<\vec{b}_2, Y\>} = q_2\; e^{i Y_2}, , & \rho^2<0;  \\
q_4\;e^{i \<\vec{b}_4, Y\>} = q_4\; e^{-i Y_2} , &  \rho^2>0.
 \end{array}
\right.
\ee
The cone crossing procedure gives us the  relations between pairs of germs 
\be
\begin{split}
({\bf Q}_{\mathbb{P}^1 \times \mathbb{P}^1} + z {\bf G}_-) (-i\psi_\Phi^2) &=  -i \p_{Y_2} W^{\mathbb{P}^1 \times \mathbb{P}^1}  = q_2\; e^{i Y_2} -  q_4\; e^{-iY_2}, \\
 ({\bf Q}_{\mathbb{P}^1 \times \mathbb{P}^1} + z {\bf G}_-)(-i\psi_\Phi^1) &=  -i \p_{Y_1} W^{\mathbb{P}^1 \times \mathbb{P}^1} = q_1\; e^{i Y_1} -  q_3\; e^{-i Y_1}.
\end{split}
\ee
Indeed we can see that the holomorphic germs belong to  the  two classes 
\be
\begin{split}
\Phi_{v} &=q_1\; e^{i Y_1}  = q_2\; e^{-i Y_1}\in H^\ast ({\bf Q}_{\mathbb{P}^1 \times \mathbb{P}^1} + z {\bf G}_-), \\
\Phi_{h}&= q_3\; e^{i Y_2} = q_4\; e^{-i Y_2}\in H^\ast ({\bf Q}_{\mathbb{P}^1 \times \mathbb{P}^1} + z {\bf G}_-).
\end{split}
\ee
We  can deform the mirror  superpotential by the holomorphic herms of the  vertical and horizontal stars, i.e. 
\be
W_{\mathbb{P}^1 \times \mathbb{P}^1} \to W_{\mathbb{P}^1 \times \mathbb{P}^1} + \e_v \Phi_{v}+ \e_h \Phi_{h} +\cO(\e^2).
\ee
The  choice  holomorphic germs gives us four different deformations of toric moduli
\be
\begin{split}
\rho^1<0, \rho^2<0  \;:\; (q_1, q_2, q_3, q_4) \to  ((1+\e_v)q_1, (1+\e_h)q_2, q_3, q_4  ),  \\ 
\rho^1>0, \rho^2<0 \;:\; (q_1, q_2, q_3, q_4) \to  (q_1, (1+\e_h)q_2, (1+\e_v)q_3, q_4  ),  \\ 
\rho^1<0, \rho^2>0 \;:\; (q_1, q_2, q_3, q_4) \to  ((1+\e_v)q_1, q_2, q_3, (1+\e_h)q_4  ),  \\ 
\rho^1>0, \rho^2>0 \;:\; (q_1, q_2, q_3, q_4) \to  (q_1, q_2, (1+\e_v)q_3, (1+\e_h)q_4  ).
\end{split}
\ee
The four  deformations above describe  the same   Kahler moduli deformation. The degree vector $\beta$ is two-dimensional and we will parametrize it by $\beta = (d_v, d_h)$.  The  star basis  representative for  the degree , i.e.  $S_\beta =  d_v H_v + d_h H_h$.  The weight factor evaluates into 
\be
\begin{split}
q^\beta &= (q_1q_3)^{d_h} ( q_2 q_4)^{d_v}  = (q_1 q_3 (1+\e_v))^{d_h} ( q_2 q_4(1+\e_h))^{d_v} \\
& = (q_1q_3)^{d_h} ( q_2 q_4)^{d_v} (1+d_h\e_v+d_v\e_h +\cO(\e^2)).
\end{split}
\ee
The $\cO(\e)$ terms in the last equality match with  intersection of degree star $S_\beta$  with a star $S = \e_v S_v + \e_h S_h$. Indeed we can evaluate 
\be
\beta \cdot S = d_v \e_v \; S_v\cdot S_v + (d_h \e_v +d_v \e_h) \; S_h \cdot S_v +d_h \e_h\; S_h \cdot S_h   = d_h\e_v+d_v\e_h.
\ee
The intersection numbers for the star observables  $S_v$ and $S_h$ are 
\be
\begin{split}
S_v \cdot S_v& = \sum_{\vec{l}, \vec{l}' \in H_v} |\vec{l}\times \vec{l}'| \chi_{\vec{l}, -\vec{l}'}(\rho) =0,\\
S_h \cdot S_h &= \sum_{\vec{l}, \vec{l}' \in H_h} |\vec{l}\times \vec{l}'| \chi_{\vec{l}, -\vec{l}'}(\rho) =0,\\
S_v \cdot S_h &= \sum_{\vec{l}\in H_v,\;\;\; \vec{l}' \in H_h} |\vec{l}\times \vec{l}'| \chi_{\vec{l}, -\vec{l}'}(\rho) =1.
\end{split}
\ee
The holomorphic germ for the point observable 
\be
\begin{split}
\Phi_\rho&=  q_1 q_2\; \chi_{\vec{b}_1, \vec{b}_2} (-\vec{\rho}\;)\;e^{i Y_1+iY_2} +q_1 q_4\; \chi_{\vec{b}_1, \vec{b}_4} (-\vec{\rho}\;)\; e^{i Y_1-iY_2}  \\
&\qquad+q_2 q_3 \; \chi_{\vec{b}_3, \vec{b}_2} (-\vec{\rho}\;)\; e^{-i Y_1+iY_2} + q_3q_4 \; \chi_{\vec{b}_2, \vec{b}_4}(-\vec{\rho}\;) \; e^{-i Y_1-iY_2}. 
\end{split}
\ee 
 Let us  provide the exact term for pairs
\be\notag
\begin{split}
({\bf Q}_{\mathbb{P}^1 \times \mathbb{P}^1} + z {\bf G}_-)(- i e^{\pm i Y_1}\psi_\Phi^2) & = -i e^{\pm i Y_1} \p_{Y_2} W_{\mathbb{P}^1 \times \mathbb{P}^1}   = q_2 e^{\pm i Y_1+i Y_2} -  q_4 e^{\pm i Y_1-iY_2}  , \\
 ({\bf Q}_{\mathbb{P}^1 \times \mathbb{P}^1} + z {\bf G}_-) (-ie^{\pm i Y_2}\psi_\Phi^1) &= -i e^{\pm i Y_2} \p_{Y_1} W_{\mathbb{P}^1 \times \mathbb{P}^1} =  q_1 e^{i Y_1\pm i Y_2} -  q_3 e^{-i Y_1\pm iY_2}.
\end{split}
\ee
Using lemma  we conclude that holomorphic germ can be written in the form 
\be\notag
\Phi_\rho = q_1 q_2 \;e^{i Y_1+iY_2} = q_2 q_3\;e^{-i Y_1+iY_2} =  q_1 q_4\; e^{i Y_1-iY_2}  =  q_3 q_4\;e^{-i Y_1-iY_2}\in H^\ast ({\bf Q}_{\mathbb{P}^1 \times \mathbb{P}^1} + z {\bf G}_-).
\ee
Tropical good section
\be
\hbox{Im}\; S^{trop}_{\mathbb{P}^1 \times \mathbb{P}^1} = \mathbb{C}\<1, \Phi_{v},  \Phi_{h}, \Phi_\rho\> = \mathbb{C}\<1, q_1 \; e^{iY_1}, q_2\;  e^{i Y_2}, q_1 q_2\; e^{iY_1+i Y_2}\>.
\ee

 \subsection{Blow up of a point on  $\mathbb{P}^2$} \label{sec_p2_bl_mirr}
 
We can depict the blow-up of  a  point  on $\mathbb{P}^2$ by cutting a corner on compactifying polyhedron  for $\mathbb{P}^2$. Similarly the corresponding fan is a refinement of the fan for $\mathbb{P}^2$.
 \newline\newline
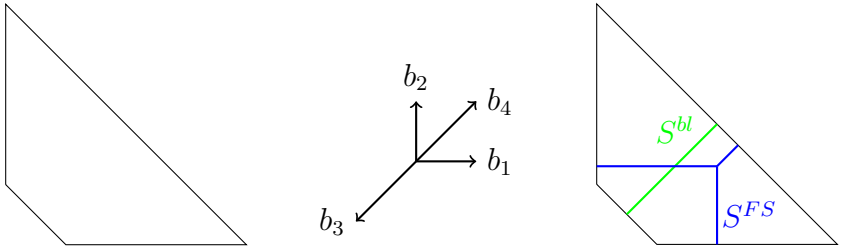
\begin{tikzpicture}[scale=0.8]
\draw (0, 1) -- (0, 4)--(4,0)--(1,0)--(0,1);
\end{tikzpicture}\qquad
\begin{tikzpicture}[scale=0.4]
\draw[color=black, thick,->](0, 0)-- (2, 0);
\draw[color=black, thick, ->] (0, 0)-- (0,2);
\draw[color=black, thick, ->](0, 0)-- (-2,-2);
\draw[color=black, thick, ->](0, 0)-- (2,2);
\node  at (2,0) [anchor=west]{$b_1$};
\node  at (0,2) [anchor=south]{$b_2$};
\node  at (-2,-2) [anchor=east]{$b_3$};
\node  at (2,2) [anchor=west]{$b_4$};
\end{tikzpicture}
\qquad
\begin{tikzpicture}[scale=0.8]
\draw (0, 1) -- (0, 4)--(4,0)--(1,0)--(0,1);
\draw[color=green, thick] (0.5, 0.5) -- (2, 2);
\draw[color=blue, thick] (0, 1.3) -- (2, 1.3);
\draw[color=blue, thick] (2, 0) -- (2, 1.3);
\draw[color=blue, thick] (2.35, 1.65) -- (2, 1.3);
\node[color=green]  at (0.8,1.9) [anchor=west]{$S^{bl}$};
\node[color=blue]  at (1.9,0.5) [anchor=west]{$S^{FS}$};
\end{tikzpicture}
\newline\newline
The compactifying divisors  for $\widehat{\mathbb{P}^2}$ are 
\be
B_{\widehat{\mathbb{P}^2}} = \{ \vec{b}_1= (1,0), \vec{b}_2=(0,1),\vec{b}_3= (-1,-1), \vec{b}_4= (1,1)\}.
\ee
The mirror superpotential (\ref{mirr_superpotential}) is  
\be
W_{\widehat{\mathbb{P}^2}} =  q_1\; e^{i Y_1} +q_2\; e^{i Y_2}  + q_3 \;e^{-i Y_1-iY_2} + q_4\;  e^{i Y_1+i Y_2}.
\ee
  The size of  $\mathbb{P}^1$ at blow up point  is controlled by $q_4$. The limit $q_4 \to 0 $ describes a blow down of $\widehat{\mathbb{P}^2}$ to $\mathbb{P}^2$, while the superpotential  in the limit becomes the mirror superpotential for $\mathbb{P}^2$.

Second Betti number $\dim H^2 (\widehat{\mathbb{P}^2})=2$, hence there are two independent hypersurface  observables. We can  choose  a basis consisting of Fubini-Study star  $S^{FS}$ on $\mathbb{P}^2$ (depicted in blue) and a two ray star  $S^{bl}$, related to the blow up,  depicted in green.
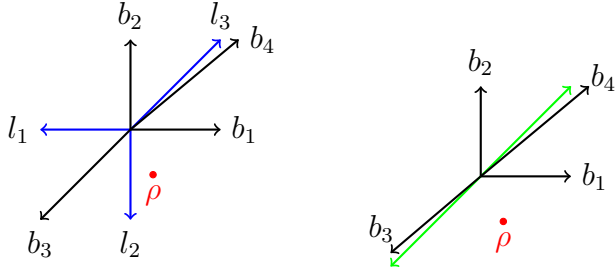
\newline\newline
\begin{tikzpicture}[scale=0.6]
\draw[color=blue, thick,->](0, 0)-- (-2, 0);
\draw[color=blue, thick, ->] (0, 0)-- (0,-2);
\draw[color=blue, thick, ->](0, 0)-- (2,2);
\node  at (-2,0) [anchor=east]{$l_1$};
\node  at (0,-2) [anchor=north]{$l_2$};
\node  at (2,2) [anchor=south]{$l_3$};
\draw[color=black, thick,->](0, 0)-- (2, 0);
\draw[color=black, thick, ->] (0, 0)-- (0,2);
\draw[color=black, thick, ->](0, 0)-- (-2,-2);
\draw[color=black, thick, ->](0, 0)-- (2.4,2);
\node  at (2,0) [anchor=west]{$b_1$};
\node  at (0,2) [anchor=south]{$b_2$};
\node  at (-2,-2) [anchor=north]{$b_3$};
\node  at (2.4,2) [anchor=west]{$b_4$};
\filldraw[red] (0.5 ,-1) circle (2pt)node[anchor=north]{$\rho$};
\end{tikzpicture}\qquad\
\begin{tikzpicture}[scale=0.6]
\draw[color=green, thick,<->](-2, -2)-- (2, 2);
\draw[color=black, thick,->](0, 0)-- (2, 0);
\draw[color=black, thick, ->] (0, 0)-- (0,2);
\draw[color=black, thick, ->](0, 0)-- (-2,-1.7);
\node  at (2,0) [anchor=west]{$b_1$};
\node  at (0,2) [anchor=south]{$b_2$};
\node  at (-2.2,-1.7) [anchor=south]{$b_3$};
\draw[color=black, thick, ->](0, 0)-- (2.4,2);
\node  at (2.2,2.2) [anchor=west]{$b_4$};
\filldraw[red] (0.5 ,-1) circle (2pt)node[anchor=north]{$\rho$};
\end{tikzpicture}
\newline
The  holomorphic germ  for Fubini-Study star   $S^{FS}_\rho$ observable 
\be
\Phi_{S^{FS}_{-\rho}}  = \left\{
\begin{array}{ll}
  q_3\; e^{i \<\vec{b}_3, Y\>} = q_3\; e^{-i Y_1-iY_2}, & \vec{\rho} \in \hbox{Cone}(\vec{l}_1, \vec{l}_2)  \\
 q_1\; e^{i \<\vec{b}_1, Y\>}  + q_4\; e^{i \<\vec{b}_4, Y\>} = q_1\; e^{i Y_1}+ q_4\; e^{i Y_1+iY_2} , & \vec{\rho} \in \hbox{Cone}(\vec{l}_2, \vec{l}_3) \\
 q_2\; e^{i \<\vec{b}_2, Y\>}+ q_4\; e^{i \<\vec{b}_4, Y\>}  = q_2\; e^{i Y_2}+ q_4\; e^{i Y_1+iY_2}, & \vec{\rho} \in \hbox{Cone}(\vec{l}_1, \vec{l}_3)
 \end{array}
\right.
\ee
The  holomorphic germ  for $S^{bl}$-observable 
\be
\Phi_{S^{bl}_{-\rho}}  = \left\{
\begin{array}{ll}
 q_1\; e^{i \<\vec{b}_1, Y\>} = q_1 e^{i Y_1}, & \rho^2>\rho^1 \\
 q_2\; e^{i \<\vec{b}_2, Y\>} = q_2 e^{i Y_2}, & \rho^2<\rho^1 
 \end{array}
\right.
\ee
The cone crossing relations are 
\be
\begin{split}
 ({\bf Q}_{\widehat{\mathbb{P}^2}} + z {\bf G}_-) (-i\psi^1) & = -i\p_1 W_{\widehat{\mathbb{P}^2}} = q_1 e^{i Y_1} - q_3 e^{-i Y_1-iY_2} + q_4 e^{i Y_1+i Y_2}\\
 ({\bf Q}_{\widehat{\mathbb{P}^2}} + z {\bf G}_-) (-i\psi^2)  &= -i\p_2W_{\widehat{\mathbb{P}^2}} = q_2  e^{i Y_2}  - q_3 e^{-i Y_1-iY_2} +q_4 e^{i Y_1+i Y_2} 
\end{split}
\ee
Hence we can express the holomorphic germs for the line observables in the form 
\be
\begin{split}
\Phi_{S^{FS}} =& q_3 e^{-i Y_1- iY_2}  = q_1 e^{i Y_1} +q_4 e^{i Y_1+i Y_2} =q_2 e^{i Y_2} + q_4 e^{i Y_1+i Y_2},\\
\Phi_{S^{bl}} =& q_1 e^{i Y_1} =   q_2 e^{i Y_2} .
\end{split}
\ee
We  can deform the mirror  superpotential by the holomorphic herms of the  vertical and horizontal stars, i.e. 
\be
W_{\widehat{\mathbb{P}^2}} \to W_{\widehat{\mathbb{P}^2}} + \e\; \Phi_{S^{FS}}+ \e_{bl}\; \Phi_{S^{bl}} +\cO(\e^2).
\ee
Hence we have  six possible deformations of toric moduli  depending on the choice of holomorphic germs
\be
\begin{split}
(q_1, q_2, q_3, q_4)& \to  (q_1(1+\e_{bl}), q_2, q_3(1+\e), q_4  ),  \\ 
(q_1, q_2, q_3, q_4) &\to  (q_1(1+\e_{bl})(1+\e), q_2, q_3, q_4(1+\e) ), \\
(q_1, q_2, q_3, q_4) &\to  (q_1(1+\e_{bl}), q_2(1+\e), q_3, q_4(1+\e) ),  \\
(q_1, q_2, q_3, q_4) &\to  (q_1, q_2(1+\e_{bl}), q_3(1+\e), q_4 ),\\
(q_1, q_2, q_3, q_4) &\to  (q_1, q_2(1+\e_{bl})(1+\e), q_3, q_4(1+\e) ),\\
(q_1, q_2, q_3, q_4) &\to  (q_1(1+\e), q_2, q_3(1+\e_{bl}), q_4(1+\e) ).
\end{split}
\ee
The four  deformations above describe  the same   Kahler moduli deformation. The degree vector $\beta$ is two-dimensional and we will parametrize it by $\beta = (d, d_{bl})$.  The  star basis  representative for  the degree , i.e.  $S_\beta =  d \;S^{FS} + d_{bl}\; S^{bl}$.  The weight factor evaluates into 
\be
\begin{split}
q^\beta &= (q_1q_2q_3)^{d} ( q_3 q_4)^{d_{bl}}  = (q_1 q_2q_3 (1+\e) (1+\e_{bl}))^{d} ( q_3 q_4(1+\e) )^{d_{bl}} \\
& = (q_1q_2 q_3)^{d} ( q_3 q_4)^{d_{bl}} (1+d\cdot \e +d_{bl} \cdot\e +d\cdot \e_{bl} +\cO(\e^2))
\end{split}
\ee
The $\cO(\e)$ terms in the last equality match with  intersection of degree star $S_\beta$  with a star observable  $S = \e \; S^{FS} + \e_{bl} \;S^{bl}$. Indeed we can evaluate 
\be
\begin{split}
S_\beta \cdot S &= d\cdot  \e \;  S^{FS} \cdot S^{FS} +(d \cdot \e_{bl} +d_{bl}\cdot  \e)  S^{FS} \cdot S^{bl} +d_{bl} \cdot \e_{bl} \; S^{bl} \cdot S^{bl} \\
&= d\cdot \e +d_{bl}\cdot  \e +d \cdot \e_{bl}.
\end{split}
\ee
The intersection numbers for the star observables  $S_v$ and $S_h$ are 
\be
\begin{split}
S^{bl} \cdot S^{bl}& = \sum_{\vec{l}, \vec{l}' \in S^{bl}} |\vec{l}\times \vec{l}'| \chi_{\vec{l}, -\vec{l}'}(\rho) =0,\\
S^{FS} \cdot S^{FS} &= \sum_{\vec{l}, \vec{l}' \in S^{FS}} |\vec{l}\times \vec{l}'| \chi_{\vec{l}, -\vec{l}'}(\rho) =1,\\
S^{FS} \cdot S^{bl} &= \sum_{\vec{l}\in S^{FS},\;\;\; \vec{l}' \in S^{bl}} |\vec{l}\times \vec{l}'| \chi_{\vec{l}, -\vec{l}'}(\rho) =1.
\end{split}
\ee
The holomorphic germ  for  the point observable at point $\vec{\rho}$,  labeled by four cones
 \be
\Phi_{\rho }  = \left\{
\begin{array}{ll}
q_1 q_2\; e^{i Y_1+iY_2} +q_2 q_4\; e^{i Y_1+2i Y_2} , & \vec{\rho} \in \hbox{Cone}(\vec{b}_2, \vec{b}_4);  \\
q_1 q_2\; e^{i Y_1+iY_2} +q_1q_4 \;e^{2i Y_1+i Y_2}  , & \vec{\rho} \in \hbox{Cone}(\vec{b}_1, \vec{b}_4); \\
q_1 q_3\; e^{-iY_2} , & \vec{\rho} \in \hbox{Cone}(\vec{b}_1, \vec{b}_3);\\
q_2 q_3\;  e^{-iY_1} , & \vec{\rho} \in \hbox{Cone}(\vec{b}_2, \vec{b}_3).
 \end{array}
\right.
\ee
 The tropical good section
\be
\hbox{Im}\; S^{trop}_{\widehat{\mathbb{P}^2}}  = \mathbb{C}\<1, \Phi_\rho,  \Phi_{S^{FS}},  \Phi_{S^{bl}}\> = \mathbb{C}\<1, q_2 q_3\; e^{ -i Y_1} , q_3 \;e^{-i Y_1- iY_2}, q_1\; e^{i Y_1}  \>.
\ee

\section{Recursion for point observables}

The holomorphic germs for hypersurface observables and point observables are quite similar. Both are linear combinations  finitely-many factors $e^{i \<\vec{m},  Y\>}$ with minor difference:  In case of line observables vectors $\vec{m} = \vec{b}$ belong to the fan $B_X$ of $X$, while in case of point observable $\vec{m} =\vec{b}+\vec{b}'$ is the sum of two vectors $\vec{b}, \vec{b}' \in B_X$ from the fan of $X$.  The deformation of the superpotential by such holomorphic germs 
\be
W_X  \to   W_X^\e = W_X + \e \Phi_P = \sum_{\vec{b}\in B_X} q_{\vec{b}}\; e^{i \<\vec{b}, Y\>} +    \sum_{\vec{b}, \vec{b}' \in B_X} c_{\vec{b}\vec{b}'} \; e^{i \<\vec{b}+\vec{b}', Y\>}
\ee
in some cases can be thought as a superpotential for  different toric variety $X_\e$, defined by the extension of the fan   $B_X$ by vectors $\vec{b}+\vec{b}'$ for each non-zero $c_{\vec{b}\vec{b}'}$. 

An extension of the fan by sum of two vectors in some cases describe a blow up of a point in a toric variety. The simplest example of such phenomenon is the blow up of a point on $\mathbb{P}^2$. Indeed the fan  for  $\widehat{\mathbb{P}^2}$
is  an extension of the fan for  $\mathbb{P}^2$ by adding a vector $\vec{b}_4 = \vec{b}_1+\vec{b}_2$ as shown on picture below
 \newline\newline
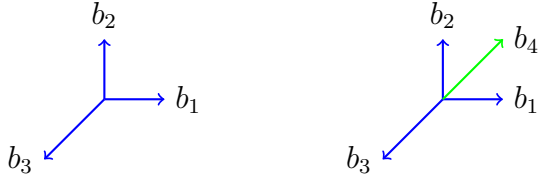
\begin{tikzpicture}[scale=0.4]
\draw[color=blue, thick,->](0, 0)-- (2, 0);
\draw[color=blue, thick, ->] (0, 0)-- (0,2);
\draw[color=blue, thick, ->](0, 0)-- (-2,-2);
\node  at (2,0) [anchor=west]{$b_1$};
\node  at (0,2) [anchor=south]{$b_2$};
\node  at (-2,-2) [anchor=east]{$b_3$};
\end{tikzpicture}\qquad\qquad
\begin{tikzpicture}[scale=0.4]
\draw[color=blue, thick,->](0, 0)-- (2, 0);
\draw[color=blue, thick, ->] (0, 0)-- (0,2);
\draw[color=blue, thick, ->](0, 0)-- (-2,-2);
\draw[color=green, thick, ->](0, 0)-- (2,2);
\node  at (2,0) [anchor=west]{$b_1$};
\node  at (0,2) [anchor=south]{$b_2$};
\node  at (-2,-2) [anchor=east]{$b_3$};
\node  at (2,2) [anchor=west]{$b_4$};
\end{tikzpicture}
\newline\newline
In the rest of this section we provide an explicit example of the superpotential deformation by the holomorphic germ for the point observable and discuss a potential implications for the tropical Gromov-Witten invariants.

\subsection{Recursion for point  observables on $\mathbb{P}^2$}

Let us consider a 4-point tropical Gromov-Witten invariant: the number of tropical curves (of degree-1 and genus-0) passing through 2 distinct points $P_1, P_2$ and two hypersurfaces $H_3, H_4$ in $\mathbb{P}^2$. 
We can use the divisor relation  to express  the 4-point Gromov-Witten invariant via  3- and 2- point invariants
\be\label{div_p2_4pt}
 \<\g_{P_1},\g_{P_2},\g_{H_3},\g_{H_4}\>^{\mathbb{P}^2}_{d=1} =1\cdot  \<\g_{P_1},\g_{P_2},\g_{H_3}\>^{\mathbb{P}^2}_{d=1}  = 1\cdot  \<\g_{P_1},\g_{P_2}\>^{\mathbb{P}^2}_{d=1}=1 .
\ee
Below we provide the enumerative proof of the relation  (\ref{div_p2_4pt}). The  tropical  hypersufraces $H_3$  and $H_4$  are 3-valent stars depicted in green and black.   From the picture we observe that both stars 
always intersect the tropical curve of degree-1, depicted in blue, at a single point.  Hence the 4-point invariant is determined by the tropical curves  passing through points $P_1, P_2$ and there is only one such curve.   
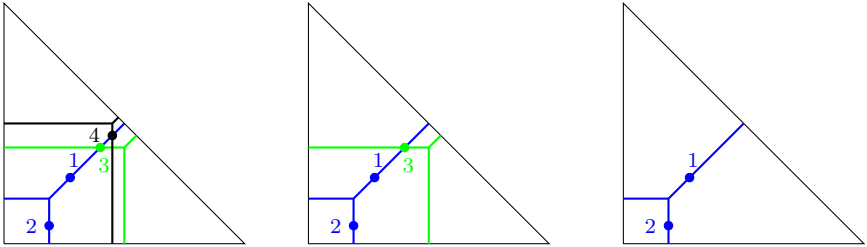
\newline\newline
\begin{tikzpicture}[scale=0.8]
\draw (0,0)-- (0, 4)--(4,0)--(0,0);
\draw[color=blue, thick] (0.75, 0.75) -- (2, 2);
\draw[color=blue, thick] (0.75, 0) -- (0.75, 0.75);
\draw[color=blue, thick] (0.75, 0.75) -- (0, 0.75) ;
\filldraw[blue] (1.1,1.1) circle (2pt) node[anchor=south]{$\;_1$};
\filldraw[blue] (0.75,0.3) circle (2pt) node[anchor=east]{$\;_2$};
\draw[color=green, thick] (0, 1.6) -- (2, 1.6) ;
\draw[color=green, thick] (2, 0) -- (2, 1.6) ;
\draw[color=green, thick]  (2, 1.6)--(2+0.2, 1.6+0.2) ;
\draw[color=black, thick] (1.8, 0) -- ( 1.8, 2) ;
\draw[color=black, thick] (0, 2) -- ( 1.8, 2) ;
\draw[color=black, thick] (1.8, 2) -- ( 1.8+0.1, 2+0.1) ;
\filldraw[black] (1.8,1.8) circle (2pt) node[anchor=east]{$\;_4$};
\filldraw[green] (1.6,1.6) circle (2pt) node[anchor=north]{$\;_3$};
\end{tikzpicture}\qquad
\begin{tikzpicture}[scale=0.8]
\draw (0,0)-- (0, 4)--(4,0)--(0,0);
\draw[color=blue, thick] (0.75, 0.75) -- (2, 2);
\draw[color=blue, thick] (0.75, 0) -- (0.75, 0.75);
\draw[color=blue, thick] (0.75, 0.75) -- (0, 0.75) ;
\filldraw[blue] (1.1,1.1) circle (2pt) node[anchor=south]{$\;_1$};
\filldraw[blue] (0.75,0.3) circle (2pt) node[anchor=east]{$\;_2$};
\filldraw[green] (1.6,1.6) circle (2pt) node[anchor=north]{$\;_3$};
\draw[color=green, thick] (0, 1.6) -- (2, 1.6) ;
\draw[color=green, thick] (2, 0) -- (2, 1.6) ;
\draw[color=green, thick]  (2, 1.6)--(2+0.2, 1.6+0.2) ;
\end{tikzpicture}
\qquad
\begin{tikzpicture}[scale=0.8]
\draw (0,0)-- (0, 4)--(4,0)--(0,0);
\draw[color=blue, thick] (0.75, 0.75) -- (2, 2);
\draw[color=blue, thick] (0.75, 0) -- (0.75, 0.75);
\draw[color=blue, thick] (0.75, 0.75) -- (0, 0.75) ;
\filldraw[blue] (1.1,1.1) circle (2pt) node[anchor=south]{$\;_1$};
\filldraw[blue] (0.75,0.3) circle (2pt) node[anchor=east]{$\;_2$};
\end{tikzpicture}
\newline\newline
We can express the 4-point Gromov-Witten  invariant via   B-model correlation function 
\be
\<\g_{P_1},\g_{P_2},\g_{H_3},\g_{H_4}\>^{\mathbb{P}^2}  =q_1 q_2 q_3\cdot  \<\g_{P_1},\g_{P_2},\g_{H_3},\g_{H_4}\>^{\mathbb{P}^2}_{d=1}  =   \<\Psi_1, \Psi_2, \Psi_3, \Psi_4\>_{Q_W}
\ee
of four mirror states 
\be
\Psi_1 = \Psi_{P_1}^W,\;\;  \Psi_2 = \Psi_{P_2}^W,\;\; \Psi_3 = \Psi_{H_3}^W,\;\; \Psi_4 = \Psi_{H_4}^W.
\ee
We can use the invariance of B-model correlation functions discussed in  \cite{Losev:2023bhj} 
\be
   \<\Psi_1, \Psi_2, \Psi_3, \Psi_4 +Q_W \chi\>_{Q_W} =  \<\Psi_1, \Psi_2, \Psi_3, \Psi_4 \>_{Q_W}
   \ee
to  replace  the  mirror state $\Psi_2$ by its holomorphic germ. In particular, let us choose  the holomorphic germ in the form 
\be
\Phi_2 =  q_1 q_2\; e^{iY_1+iY_2},
\ee
to rewrite the B-model correlation function in the following form
\be
 \<\Psi_1, \Psi_2, \Psi_3, \Psi_4\>_{Q_W} = \<\Psi_1, \Phi_2, \Psi_3, \Psi_4\>_{Q_W}.
\ee
We can use the  recursion formula  from  \cite{Losev:2023bhj}   to express the  4-point  function as a derivative of 3-point function
\be\label{4_to_3_recursion}
\<\Psi_1, \Phi_2, \Psi_3, \Psi_4\>_{Q_W} = \frac{d}{d\e}\Big|_{\e=0} \<\Psi^\e_1, \Psi^\e_3, \Psi^\e_4\>_{Q_{W^\e}}
\ee
in B-model with    deformed  superpotential 
\be\label{p2_deform_superpt}
W_{\mathbb{P}^2}^\e  =W_{\mathbb{P}^2} + \e\; \Phi_2=  q_1\; e^{iY_1} + q_2\; e^{iY_2} + q_3\; e^{- iY_1-iY_2} + \e q_1 q_2\; e^{iY_1+iY_2} = W_{X_\e}.
\ee
The deformed superpotential is the mirror superpotential for the different  toric manifold $X_\e =\widehat{\mathbb{P}^2}$.  The  polytopes for $\mathbb{P}^2$  and $X_\e$  are depicted below
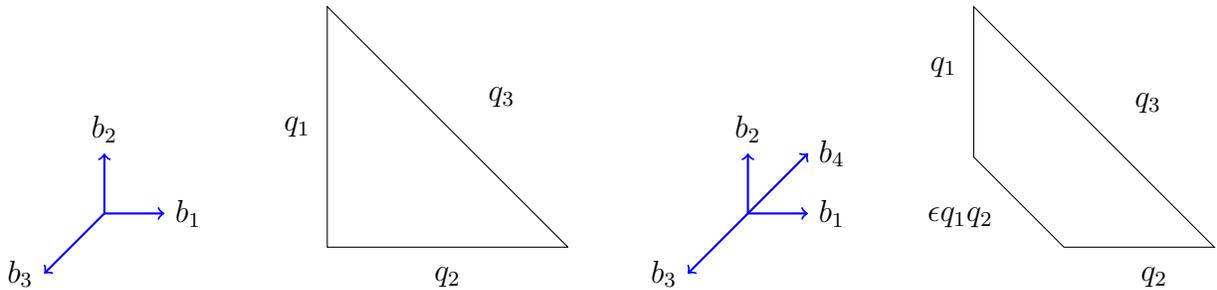
\newline\newline
\begin{tikzpicture}[scale=0.4]
\draw[color=blue, thick,->](0, 0)-- (2, 0);
\draw[color=blue, thick, ->] (0, 0)-- (0,2);
\draw[color=blue, thick, ->](0, 0)-- (-2,-2);
\node  at (2,0) [anchor=west]{$b_1$};
\node  at (0,2) [anchor=south]{$b_2$};
\node  at (-2,-2) [anchor=east]{$b_3$};
\end{tikzpicture}\qquad
\begin{tikzpicture}[scale=0.8]
\draw(4,0)-- (0, 0) -- (0, 4)--(4,0);
\node  at (2.5,2.5)[anchor=west]  {$q_3$};
\node  at (2,-0.5) {$q_2$};
\node  at (-0.5 ,2) {$q_1$};
\end{tikzpicture}
\qquad
\begin{tikzpicture}[scale=0.4]
\draw[color=blue, thick,->](0, 0)-- (2, 0);
\draw[color=blue, thick, ->] (0, 0)-- (0,2);
\draw[color=blue, thick, ->](0, 0)-- (-2,-2);
\draw[color=blue, thick, ->](0, 0)-- (2,2);
\node  at (2,0) [anchor=west]{$b_1$};
\node  at (0,2) [anchor=south]{$b_2$};
\node  at (-2,-2) [anchor=east]{$b_3$};
\node  at (2,2) [anchor=west]{$b_4$};
\end{tikzpicture}\qquad
\begin{tikzpicture}[scale=0.8]
\draw (0, 1.5) -- (0, 4)--(4,0)--(1.5,0)--(0,1.5);
\node  at (2.5,2.4)[anchor=west] {$q_3$};
\node  at (3,-0.5) {$q_2$};
\node  at (-0.5 ,3){$q_1$};
\node  at (0.5 ,0.5)[anchor=east] {$\e q_1q_2$};
\end{tikzpicture}
\newline\newline
We showed in \cite{Losev:2023bhj} that the  deformed mirror  states in correlation function (\ref{4_to_3_recursion})  are mirror states in deformed theory, i.e.
\be\label{state_deform_qm}
\Psi_\a^\e  = \Psi^W_\a  + 2\pi K_{W} G_- \mu_2 (\Psi^W_\a ,\e \; \Phi_2) =  \Psi_\a^{W^\e} .
\ee
Hence we can  represent   the 3-point function  $ \<\Psi^\e_1, \Psi^\e_3, \Psi^\e_4\>_{Q_{W_\e}}$  as  the  sum of A-model amplitudes for  three observables $P_1, H_3, H_4$ in the HTQM for $X^\e = \widehat{\mathbb{P}^2}$ and then  convert them into tropical Gromov-Witten invariants on $X^\e = \widehat{\mathbb{P}^2}$. Namely
\be
 \<\Psi_{P_1}^{\e}, \Psi_{H_3}^{\e}, \Psi_{H_4}^{\e}\>_{Q_{W^\e}}  = \< \g_{ P_1}, \g_{H_3}, \g_{H_4} \>^{ \widehat{\mathbb{P}^2}}. 
\ee
$X_\e= \widehat{\mathbb{P}^2}$ is a toric space, hence the Gromov-Witten invariant is a polynomial in toric moduli $q_1,q_2, q_3$ and new module $q_4(\e) = \e q_1q_2$.     The $\mathbb{P}^2$-invariants   are polynomials in Kahler module  $q= q_1q_2q_3$. 
  
 The $\widehat{\mathbb{P}^2} $-invariants   are polynomials in Kahler moduli $q, q_{bl}$  where $q_{bl} = q_4 q_3$ in additional  Kahler module on $\widehat{\mathbb{P}^2}$  . Hence we can expand
\be
\<\g_{ P_1}, \g_{H_3}, \g_{H_4}\>^{ \widehat{\mathbb{P}^2}} = \sum_{d, d_\e =0}^\infty  q^d  q_{bl}^{d_{bl}}\< \g_{ P_1}, \g_{H_3}, \g_{H_4}\>^{ \widehat{\mathbb{P}^2}}_{d, d_{bl}}
\ee
The product $q_{bl} = q_4 q_3$ is a Kahler module of $ \widehat{\mathbb{P}^2}$ associated to the size of the blow-up $\mathbb{P}^1$.  The derivative at $\e=0$ picks up monomials, linear in $\e$, hence linear in $q_4(\e)$  and Kahler module $q_{bl}$, i.e.
\be
 \frac{d}{d\e}\Big|_{\e=0}  \<\g_{ P_1}, \g_{H_3}, \g_{H_4} \>^{ \widehat{\mathbb{P}^2}} =\frac{d}{d\e}q_{bl}\cdot  \sum_{d =0}^\infty  q^d \< \g_{P_1}, \g_{H_3}, \g_{H_4}\>^{ \widehat{\mathbb{P}^2}}_{d, d_{bl}=1}  =q \< \g_{ P_1}, \g_{H_3}, \g_{H_4}\>^{ \widehat{\mathbb{P}^2}}_{0, d_{bl}=1}, 
\ee
where we used 
\be
\frac{d}{d\e}q_{bl} = q_3 \frac{d}{d\e}q_4 = q_1q_2q_3 =q
\ee
and the degree selection argument. The dimension of moduli space of tropical curves of bi-degree $(d, d_{bl})$ on  $ \widehat{\mathbb{P}^2}$ with $3$ marked points  should be equal to the total degree of three observables, which implies that 
\be
3d+2d_{bl}+2 = \sum_{\a=1}^3 \deg \g_\a = 4.
\ee 
Hence the Gromov-Witten invariant $\<\g_{ P_1}, \g_{H_3}, \g_{H_4}\>^{ \widehat{\mathbb{P}^2}}$ is non-zero only for bi-degree $d=0, d_{bl}=1$.

The result of this procedure relates the 4pt  degree-1 Gromov-Witten  invariant  on $\mathbb{P}^2$ to 3pt invariant on  $\widehat{\mathbb{P}^2}$ of bi-degree $(0,1)$, i.e 
\be\label{4_to_3_corners_p2}
\<\g_{ P_1}, \g_{P_2},\g_{H_3}, \g_{H_4}\>^{\mathbb{P}^2}_{d=1}  =  \< \g_{ P_1}, \g_{H_3}, \g_{H_4}\>^{\widehat{\mathbb{P}^2}}_{d = 0, d_{bl}=1}. 
\ee

\subsection{Enumerative description of  recursion}

We can give an enumerative interpretation of the relation (\ref{4_to_3_corners_p2}) as a {\it cutting corners procedure} for tropical Gromov-Witten invariants. In case point $P_2$ is close to the corner, formed by hyperplanes supported on $\vec{b}_1$ and $\vec{b}_2$  the 
tropical Gromov-Witten invariant $\< \g_{ P_1}, \g_{P_2},\g_{H_3}, \g_{H_4} \>^{\mathbb{P}^2}$ is  supported by the diagram below.  Let us cut the corner together with the part of tropical curve and marked point $P_2$. The result of the cutting is the 
polyhedron for $\widehat{\mathbb{P}^2}$ with a tropical curve and remaining three observables: point  $P_1$ and two hyperplanes $H_1$ and $H_2$. 
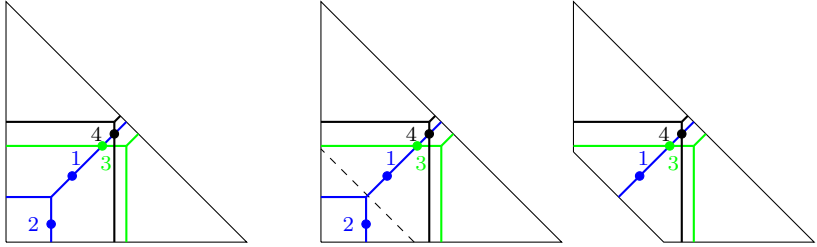
\newline\newline
\begin{tikzpicture}[scale=0.8]
\draw (0,0)-- (0, 4)--(4,0)--(0,0);
\draw[color=blue, thick] (0.75, 0.75) -- (2, 2);
\draw[color=blue, thick] (0.75, 0) -- (0.75, 0.75);
\draw[color=blue, thick] (0.75, 0.75) -- (0, 0.75) ;
\filldraw[blue] (1.1,1.1) circle (2pt) node[anchor=south]{$\;_1$};
\filldraw[blue] (0.75,0.3) circle (2pt) node[anchor=east]{$\;_2$};
\draw[color=green, thick] (0, 1.6) -- (2, 1.6) ;
\draw[color=green, thick] (2, 0) -- (2, 1.6) ;
\draw[color=green, thick]  (2, 1.6)--(2+0.2, 1.6+0.2) ;
\draw[color=black, thick] (1.8, 0) -- ( 1.8, 2) ;
\draw[color=black, thick] (0, 2) -- ( 1.8, 2) ;
\draw[color=black, thick] (1.8, 2) -- ( 1.8+0.1, 2+0.1) ;
\filldraw[black] (1.8,1.8) circle (2pt) node[anchor=east]{$\;_4$};
\filldraw[green] (1.6,1.6) circle (2pt) node[anchor=north]{$\;_3$};
\end{tikzpicture}
\qquad
\begin{tikzpicture}[scale=0.8]
\draw (0,0)-- (0, 4)--(4,0)--(0,0);
\draw[dashed]  (1.55,0)-- (0, 1.55);
\draw[color=blue, thick] (0.75, 0.75) -- (2, 2);
\draw[color=blue, thick] (0.75, 0) -- (0.75, 0.75);
\draw[color=blue, thick] (0.75, 0.75) -- (0, 0.75) ;
\filldraw[blue] (1.1,1.1) circle (2pt) node[anchor=south]{$\;_1$};
\filldraw[blue] (0.75,0.3) circle (2pt) node[anchor=east]{$\;_2$};
\draw[color=green, thick] (0, 1.6) -- (2, 1.6) ;
\draw[color=green, thick] (2, 0) -- (2, 1.6) ;
\draw[color=green, thick]  (2, 1.6)--(2+0.2, 1.6+0.2) ;
\draw[color=black, thick] (1.8, 0) -- ( 1.8, 2) ;
\draw[color=black, thick] (0, 2) -- ( 1.8, 2) ;
\draw[color=black, thick] (1.8, 2) -- ( 1.8+0.1, 2+0.1) ;
\filldraw[black] (1.8,1.8) circle (2pt) node[anchor=east]{$\;_4$};
\filldraw[green] (1.6,1.6) circle (2pt) node[anchor=north]{$\;_3$};
\end{tikzpicture}
\begin{tikzpicture}[scale=0.8]
\draw (0, 1.5) -- (0, 4)--(4,0)--(1.5,0)--(0,1.5);
\draw[color=blue, thick] (0.75, 0.75) -- (2, 2);
\filldraw[blue] (1.1,1.1) circle (2pt) node[anchor=south]{$\;_1$};
\draw[color=green, thick] (0, 1.6) -- (2, 1.6) ;
\draw[color=green, thick] (2, 0) -- (2, 1.6) ;
\draw[color=green, thick]  (2, 1.6)--(2+0.2, 1.6+0.2) ;
\draw[color=black, thick] (1.8, 0) -- ( 1.8, 2) ;
\draw[color=black, thick] (0, 2) -- ( 1.8, 2) ;
\draw[color=black, thick] (1.8, 2) -- ( 1.8+0.1, 2+0.1) ;
\filldraw[black] (1.8,1.8) circle (2pt) node[anchor=east]{$\;_4$};
\filldraw[green] (1.6,1.6) circle (2pt) node[anchor=north]{$\;_3$};
\end{tikzpicture}
\newline\newline
The remaining tropical curve is a curve of bi-degree $(d, d_{bl}) = (0,1)$. The moduli  space of such curve  is $\mathbb{R}^1 \times S^1$: the radial part corresponds to the parallel translation of the curve as shown on a picture below
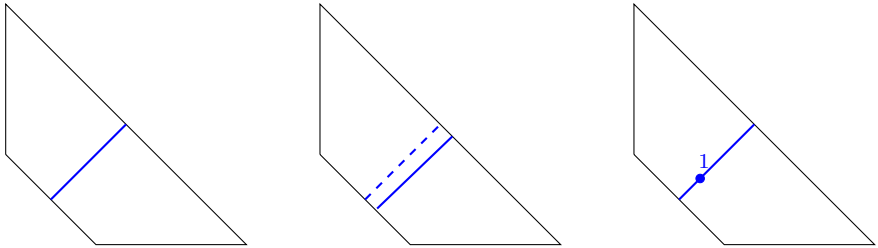
\newline\newline
\begin{tikzpicture}[scale=0.8]
\draw (0, 1.5) -- (0, 4)--(4,0)--(1.5,0)--(0,1.5);
\draw[color=blue, thick] (0.75, 0.75) -- (2, 2);
\end{tikzpicture}
\qquad
\begin{tikzpicture}[scale=0.8]
\draw (0, 1.5) -- (0, 4)--(4,0)--(1.5,0)--(0,1.5);
\draw[color=blue, thick, dashed] (0.75, 0.75) -- (2, 2);
\draw[color=blue, thick] (0.95, 0.6) -- (2.2, 1.8);
\end{tikzpicture}
\qquad
\begin{tikzpicture}[scale=0.8]
\draw (0, 1.5) -- (0, 4)--(4,0)--(1.5,0)--(0,1.5);
\draw[color=blue, thick] (0.75, 0.75) -- (2, 2);
\filldraw[blue] (1.1,1.1) circle (2pt) node[anchor=south]{$\;_1$};
\end{tikzpicture}
\newline\newline
The three remaining observables 
completely fix the moduli  hence  there exists a tropical  Gromov-Witten invariant on $ \widehat{\mathbb{P}^2}$, which   counts the number of degree-$(0,1)$  tropical curves $\G$ which pass through the point $P_1$ and two hyperplanes $H_3$ and $H_4$. 
Both hyperplanes $H_3$  and $H_4$ have degree-$(1,0)$  and intersect all curves $\G$ with  intersection numbers $H_3\cdot \G = H_4\cdot \G =1$, hence we can reduce the original problem to  counting curves $\G$ through point $P_1$. 
There is a unique such tropical curve,  hence 
\be
\< \g_{ P_1}, \g_{H_3}, \g_{H_4}\>^{\widehat{\mathbb{P}^2}}_{d = 0, d_{bl}=1}=1.
\ee

We can apply the cutting corner procedure to other tropical Gromov-Witten invariants. For example we can consider a degree-2 curves through 5 distinct points on $\mathbb{P}^2$.  Among the   4 distinct tropical curves of degree-2  let us consider the one presented below.
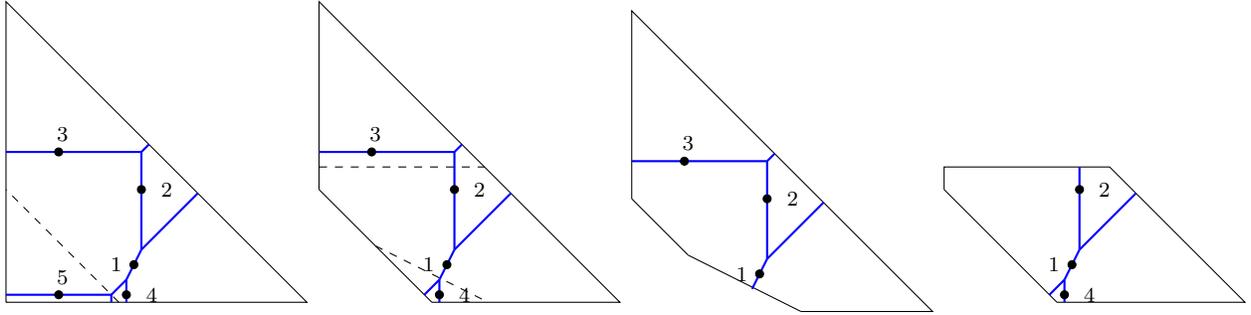
\newline\newline
\begin{tikzpicture}[scale=1]
\draw(0,0)-- (4, 0) -- (0, 4)--(0,0);
\draw [dashed] (1.5, 0) --(0,1.5);
\draw[color=blue, thick] (1.6, 0) -- (1.6, 0.3);
\draw[color=blue, thick] (1.6, 0.3) -- (1.4, 0.1);
\draw[color=blue, thick] (1.4, 0.1) -- (1.4, 0);
\draw[color=blue, thick] (0, 0.1) -- (1.4, 0.1);
\draw[color=blue, thick] (1.6, 0.3) -- (1.8, 0.7);
\draw[color=blue, thick] (1.8, 0.7) -- (1.8+0.75, 0.7+0.75);
\draw[color=blue, thick] (1.8, 0.7) -- (1.8, 2);
\draw[color=blue, thick] (0, 2) -- (1.8, 2);
\draw[color=blue, thick] (1.8, 2) -- (1.8+0.1, 2+0.1);
\filldraw[black] (0.7, 0.1) circle (1.5pt) node[anchor=south]{$\;_5$};
\filldraw[black] (0.7, 2) circle (1.5pt) node[anchor=south]{$\;_3$};
\filldraw[black] (1.6, 0.1) circle (1.5pt) node[anchor=west]{$\;_4$};
\filldraw[black] (1.8, 1.5) circle (1.5pt) node[anchor=west]{$\;_2$};
\filldraw[black] (1.7, 0.5) circle (1.5pt) node[anchor=east]{$\;_1$};
\end{tikzpicture}
\begin{tikzpicture}[scale=1]
\draw(1.5,0)-- (4, 0) -- (0, 4)--(0,1.5);
\draw (1.5, 0) --(0,1.5);
\draw[dashed](0.75,0.75)-- (2.25, 0) ;
\draw[dashed](0,1.8)-- (2.2, 1.8) ;
\draw[color=blue, thick] (1.6, 0) -- (1.6, 0.3);
\draw[color=blue, thick] (1.6, 0.3) -- (1.4, 0.1);
\draw[color=blue, thick] (1.6, 0.3) -- (1.8, 0.7);
\draw[color=blue, thick] (1.8, 0.7) -- (1.8+0.75, 0.7+0.75);
\draw[color=blue, thick] (1.8, 0.7) -- (1.8, 2);
\draw[color=blue, thick] (0, 2) -- (1.8, 2);
\draw[color=blue, thick] (1.8, 2) -- (1.8+0.1, 2+0.1);
\filldraw[black] (0.7, 2) circle (1.5pt) node[anchor=south]{$\;_3$};
\filldraw[black] (1.6, 0.1) circle (1.5pt) node[anchor=west]{$\;_4$};
\filldraw[black] (1.8, 1.5) circle (1.5pt) node[anchor=west]{$\;_2$};
\filldraw[black] (1.7, 0.5) circle (1.5pt) node[anchor=east]{$\;_1$};
\end{tikzpicture}
\begin{tikzpicture}[scale=1]
\draw(2.25,0)-- (4, 0) -- (0, 4)--(0,1.5);
\draw (0.75,0.75) --(0,1.5);
\draw (0.75,0.75)-- (2.25, 0) ;
\draw[color=blue, thick] (1.6, 0.3) -- (1.8, 0.7);
\draw[color=blue, thick] (1.8, 0.7) -- (1.8+0.75, 0.7+0.75);
\draw[color=blue, thick] (1.8, 0.7) -- (1.8, 2);
\draw[color=blue, thick] (0, 2) -- (1.8, 2);
\draw[color=blue, thick] (1.8, 2) -- (1.8+0.1, 2+0.1);
\filldraw[black] (0.7, 2) circle (1.5pt) node[anchor=south]{$\;_3$};
\filldraw[black] (1.8, 1.5) circle (1.5pt) node[anchor=west]{$\;_2$};
\filldraw[black] (1.7, 0.5) circle (1.5pt) node[anchor=east]{$\;_1$};
\end{tikzpicture}
\begin{tikzpicture}[scale=1]
\draw(1.5,0)-- (4, 0) --  (2.2, 1.8) -- (0,1.8) --(0,1.5);
\draw (1.5, 0) --(0,1.5);
\draw[color=blue, thick] (1.6, 0) -- (1.6, 0.3);
\draw[color=blue, thick] (1.6, 0.3) -- (1.4, 0.1);
\draw[color=blue, thick] (1.6, 0.3) -- (1.8, 0.7);
\draw[color=blue, thick] (1.8, 0.7) -- (1.8+0.75, 0.7+0.75);
\draw[color=blue, thick] (1.8, 0.7) -- (1.8, 1.8);
\filldraw[black] (1.6, 0.1) circle (1.5pt) node[anchor=west]{$\;_4$};
\filldraw[black] (1.8, 1.5) circle (1.5pt) node[anchor=west]{$\;_2$};
\filldraw[black] (1.7, 0.5) circle (1.5pt) node[anchor=east]{$\;_1$};
\end{tikzpicture}
\newline\newline
 We performed a single corner cut to reduce the number of marked  points to 4 and changed the target to $\widehat{\mathbb{P}^2}$.   We can continue the procedure and cut one more corner to reduce the number of points to 3. 
There are two possible cuts (up to an isomorphism) that we can perform:
\begin{itemize}
\item {\bf two points are far away}:   The resulting polytope describes the toric variety $X_{\e_1\e_2}$ which is  a blow up of $\mathbb{P}^2$ at two points. In particular  we have a network of  blow down maps
 $\pi_1, \pi_2 : X_{\e_1\e_2} \to \widehat{\mathbb{P}^2}$ which can be applied in any order. The cycles  which are pre-images of blow up   points do not intersect. 
 
\item {\bf two points are nearby}:  The resulting polyhedron describes the toric variety $X_{\e_1\e_2}$ which is  a  bi-rational transformation of  $\mathbb{P}^2$ obtained by  two consecutive  blow-ups. We have a single chain of blow  down maps. 

 \end{itemize}

\subsection{Double deformation and contact terms}

Let us give a detailed description of the geometry for two cuts  of $\widehat{\mathbb{P}^2}$.  We can use the polytopes to construct the corresponding fans
 \newline\newline
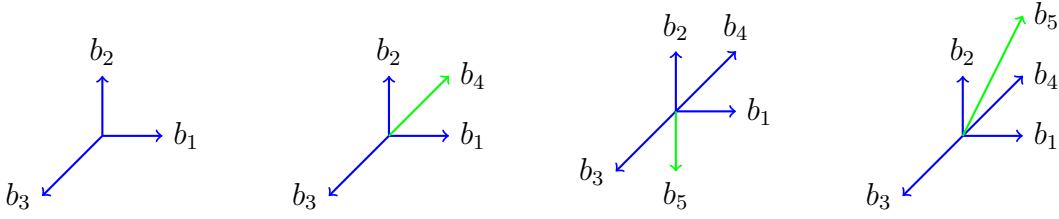
 \begin{tikzpicture}[scale=0.4]
\draw[color=blue, thick,->](0, 0)-- (2, 0);
\draw[color=blue, thick, ->] (0, 0)-- (0,2);
\draw[color=blue, thick, ->](0, 0)-- (-2,-2);
\node  at (2,0) [anchor=west]{$b_1$};
\node  at (0,2) [anchor=south]{$b_2$};
\node  at (-2,-2) [anchor=east]{$b_3$};
\end{tikzpicture}
\qquad
\begin{tikzpicture}[scale=0.4]
\draw[color=blue, thick,->](0, 0)-- (2, 0);
\draw[color=blue, thick, ->] (0, 0)-- (0,2);
\draw[color=blue, thick, ->](0, 0)-- (-2,-2);
\draw[color=green, thick, ->](0, 0)-- (2,2);
\node  at (2,0) [anchor=west]{$b_1$};
\node  at (0,2) [anchor=south]{$b_2$};
\node  at (-2,-2) [anchor=east]{$b_3$};
\node  at (2,2) [anchor=west]{$b_4$};
\end{tikzpicture}
\qquad
\begin{tikzpicture}[scale=0.4]
\draw[color=blue, thick,->](0, 0)-- (2, 0);
\draw[color=blue, thick, ->] (0, 0)-- (0,2);
\draw[color=blue, thick, ->](0, 0)-- (-2,-2);
\draw[color=blue, thick, ->](0, 0)-- (2,2);
\draw[color=green, thick, ->](0, 0)-- (0,-2);
\node  at (2,0) [anchor=west]{$b_1$};
\node  at (0,2) [anchor=south]{$b_2$};
\node  at (-2,-2) [anchor=east]{$b_3$};
\node  at (2,2) [anchor=south]{$b_4$};
\node  at (0,-2) [anchor=north]{$b_5$};
\end{tikzpicture}
\qquad
\begin{tikzpicture}[scale=0.4]
\draw[color=blue, thick,->](0, 0)-- (2, 0);
\draw[color=blue, thick, ->] (0, 0)-- (0,2);
\draw[color=blue, thick, ->](0, 0)-- (-2,-2);
\draw[color=blue, thick, ->](0, 0)-- (2,2);
\draw[color=green, thick, ->](0, 0)-- (2,4);
\node  at (2,0) [anchor=west]{$b_1$};
\node  at (0,2) [anchor=south]{$b_2$};
\node  at (-2,-2) [anchor=east]{$b_3$};
\node  at (2,2) [anchor=west]{$b_4$};
\node  at (2,4) [anchor=west]{$b_5$};
\end{tikzpicture}
\newline\newline
However in order to describe the toric moduli $q_1,...,q_5$ in terms of deformation parameters $\e_1, \e_2$  and toric moduli of the base $\mathbb{P}^2$ we need to construct the corresponding mirror superpotentials.     
Both   superpotentials  are $\e_2$-deformation of the superpotential (\ref{p2_deform_superpt}), i.e.
\be\label{consec_deform}
W_{\mathbb{P}^2}^{\e_1\e_2} = W_{\mathbb{P}^2}^{\e_1} +  \e_2 \;\Phi^{\e_1}_4.
\ee
where $\Phi^{\e_1}_4$ is the  holomorphic germ on $\widehat{\mathbb{P}^2}$.  We can express the holomorphic germs on $\widehat{\mathbb{P}^2}$ using the holomorphic germs on $\mathbb{P}^2$ then the double deformed superpotential takes the form 
\be\label{double_deform}
W_{\mathbb{P}^2}^{\e_1\e_2}  =  W_{\mathbb{P}^2} +   \e_1\; \Phi_5 +\e_2\; \Phi_4 + \e_1\e_2\; C^{trop}_W  (\Phi_4, \Phi_5).
\ee
The second equality  describes a double deformation of superpotential by a pair of holomorphic functions. The $\e_1\e_2$-term is the tropical   contact term   defined in section \ref{sec_good_section}

The two cutting corners cases correspond to the different choices of the holomorphic germs $\Phi_4, \Phi_5$ for point observable on  $\mathbb{P}^2$.  We can use our analysis from section \ref{sec_p2_bl_mirr}  for   holomorphic germs  to perform the superpotential analysis. 
\begin{itemize}
\item {\bf two points are far away}:  The holomorphic germ  for $P_4$ observable is the same for  $\mathbb{P}^2$ and $\widehat{\mathbb{P}^2}$
\be
\Phi_3^{\e_1} = \Phi_3 = q_1 q_3\; e^{-iY_2}
\ee
hence the double deformation of the  mirror superpotential is
 \be
 W_{\mathbb{P}^2}^{\e_1e_2} = W_{\mathbb{P}^2}^{\e_1} +  \e_2 \Phi_3^{\e_1} =  q_1\; e^{i Y_1} + q_2\; e^{i Y_2}  + q_3\;  e^{-i Y_1-iY_2} + \e_1 q_1 q_2 \; e^{i Y_1+i Y_2}  +\e_2  q_1 q_3\;  e^{-i Y_2}.
 \ee
There is no quadratic terms in $\e$ in our expression hence we expect that the contact term between $\Phi_3$ and $\Phi_5$ vanishes.  Indeed   the product  $\Phi_3\Phi_5$ is in image of good section (\ref{trop_good_sec_p2})
\be
\Phi_5\cdot  \Phi_3 =   q_1 q_2 \;   e^{i Y_1 +i Y_2} \cdot  q_1 q_3\; e^{- i Y_2} = q_1^2q_2q_3 \; e^{i Y_1} \in \hbox{Im} \;S_{\mathbb{P}^2}^{trop},
\ee
hence  contact term between two   deformations  is trivial, i.e.
\be
C_W^{trop} (\Phi_5, \Phi_3) = C_W^{trop} ( e^{i Y_1 +i Y_2}, e^{- i Y_2}) = 0.
\ee

 \item {\bf two points are nearby}:  The holomorphic germs
 \be
\Phi^{\e_1}_4 =q_1 q_2\; e^{i Y_1+iY_2} +q_4(\e_1) q_2\; e^{i Y_1+2i Y_2} =  q_1 q_2\; e^{i Y_1+iY_2} +\e_1 q_1 q_2^2\; e^{i Y_1+2i Y_2}
\ee
 \be
\Phi_4 = \Phi^{\e_1}_4\Big|_{\e_1=0} =  q_1 q_2\; e^{i Y_1+iY_2} 
\ee
 gives us a mirror superpotential 
  \be
 W^{\e_1\e_2}_{\mathbb{P}^2} = q_1\; e^{i Y_1} + q_2\; e^{i Y_2}  + q_3 \; e^{-i Y_1-iY_2} + (\e_1+\e_2) q_1 q_2\;  e^{i Y_1+i Y_2}  +\e_1\e_2 q_1 q_2^2\; e^{i Y_1+2i Y_2}.
 \ee
 Note that the $\e_1$ and $\e_2$ enter symmetrically.  
The quadratic term is a contact term for two (identical) deformations 
 \be
 \begin{split}
 C_W^{trop}  (\Phi_5, \Phi_4)  &= C_W^{trop} (q_1 q_2 \;e^{i Y_1 +i Y_2}, q_1q_2 \; e^{i Y_1 +i Y_2}) \\
 &= {\bf G}_- {\bf \Sigma}_W (\Phi_4 \Phi_5 - S_W \pi_W (\Phi_4\Phi_5))  ={\bf G}_- (q_1^2q_2\; e^{2i Y_1+i Y_2} i\psi_\Phi^2) \\
 & =   e^{2i Y_1+i Y_2}.
 \end{split}
 \ee
 We used 
 \be
  \pi_W (\Phi_4\Phi_5) = \pi_W (q_1^2 q_2^2\; e^{2i Y_1 +2i Y_2} ) = q_1^2q_2q_3\; e^{i Y_1}
 \ee
 and 
 \be
 \Phi_4 \Phi_5 - S_W \pi_W (\Phi_4\Phi_5) = q_1^2 q_2^2 \; e^{2i Y_1 +2i Y_2} - q_1^2q_2q_3\; e^{i Y_1} = {\bf Q}_W (q_1^2q_2 e^{2i Y_1+i Y_2} i\psi_\Phi^2). 
 \ee
\end{itemize} 
We explicitely checked  that the two ways (\ref{consec_deform}) and (\ref{double_deform}) of constructing the double deformed mirror superpotential for $\mathbb{P}^2$ give identical results when we use the  tropical good section (\ref{trop_good_sec_p2})  for the contact terms.

\subsection{Conclusion and open questions}

We described the cutting corners procedure and its application to  4- and 5- point tropical Gromov-Witten invariants on $\mathbb{P}^2$.  It is reasonable to conjecture that the cutting corners relation  (\ref{4_to_3_corners_p2}) for 4-point correlation function generalizes to  $n$-point functions
\be
\<\g_{1}, \g_{2},...,\g_{n}, \g_{P}\>^{\mathbb{P}^2}_{d}  = \<\g_{1}, \g_{2},...,\g_{n}\>^{\widehat{\mathbb{P}^2}}_{d-1, d_{bl}=1}. 
\ee
In our examples we cut up to two corners, but we can conjecture that the procedure can be iterated. If so, then we can repeat the cutting corners procedure till we are down to three point  correlation function, which we can evaluate using the residue formula in Landau-Ginzburg-Saito theory.   In particular, it would be interesting to perform the five cutting corners to evaluate    the first nontrivial Gromov-Witten  invariant, which is  12 degree-3 genus-0 curves passing through generic 8 points
on $\mathbb{P}^2$.   

There is a famous isomorphism between the blow up of two points on $\mathbb{P}^2$ and the blow up of one point on $\mathbb{P}^1\times \mathbb{P}^1$. Such  relation implies that the iterated cutting corners procedure 
after first few steps will give us the same toric spaces.  Hence, we can use this relation as consistency check of the tropical Gromov-Witten  invariants evaluation through the cutting corners procedure.

In case we can repeat the cutting corners procedure indefinitely we can use it give a non-perturbative definition of the Gromov-Witten invariants for point observables in way similar to what was done for the hyperplane observables.

\section*{Acknowledgments}

We are grateful to  Yasha Neiman   for many discussions on the topics presented in this paper.  The work  of A.L. is supported  by the Basic Research Program of the National Research University Higher School of Economics and by Wu Wen-Tsun Key Lab of Mathematics. The work of V.L. is   supported by the Quantum Gravity Unit of the Okinawa Institute of Science and Technology Graduate University (OIST).

\bibliography{surf_mirr_ref}{}

\providecommand{\href}[2]{#2}\begingroup\raggedright\begin{thebibliography}{10}

\bibitem{vafa2003mirror}
C.~Vafa and E.~Zaslow, {\em Mirror Symmetry: Clay Mathematics Monographs, Vol.
  1}.
\newblock AMS-CMI, 2003.

\bibitem{saito1983period}
K.~Saito, ``Period mapping associated to a primitive form,'' {\em Publications
  of the Research Institute for Mathematical Sciences} {\bfseries 19} no.~3,
  (1983) 1231--1264.

\bibitem{Losev:1998dv}
A.~Losev, ``{'Hodge strings' and elements of K. Saito's theory of the primitive
  form},'' in {\em {Taniguchi Symposium on Topological Field Theory, Primitive
  Forms and Related Topics}}, pp.~305--335.
\newblock 1, 1998.
\newblock \href{http://arxiv.org/abs/hep-th/9801179}{{\ttfamily
  arXiv:hep-th/9801179}}.

\bibitem{Losev:2022tzr}
A.~Losev and V.~Lysov, ``{Tropical Mirror},''
  \href{http://arxiv.org/abs/2204.06896}{{\ttfamily arXiv:2204.06896
  [hep-th]}}.

\bibitem{Losev:2023bhj}
A.~Losev and V.~Lysov, ``{Tropical Mirror Symmetry: Correlation functions},''
  \href{http://arxiv.org/abs/2301.01687}{{\ttfamily arXiv:2301.01687
  [hep-th]}}.

\bibitem{oda1988convex}
T.~Oda, ``Convex bodies and algebraic geometry. An introduction to the theory
  of toric varieties,'' {\em Ergebnisse der Mathematik und ihrer Grenzgebiete}
  {\bfseries 3} (1988) 15.

\bibitem{telen2022introduction}
S.~Telen, ``Introduction to Toric Geometry,'' 2022.

\bibitem{Mikhalkin2004}
G.~Mikhalkin, {\em Amoebas of Algebraic Varieties and Tropical Geometry},
  pp.~257--300.
\newblock Springer US, Boston, MA, 2004.

\bibitem{Mikh}
G.~Mikhalkin, ``Introduction to Tropical Geometry (notes from the IMPA lectures
  in Summer 2007),'' 2007.

\bibitem{mikhalkin2009tropical}
G.~Mikhalkin and J.~Rau, {\em Tropical geometry}, vol.~8.
\newblock MPI for Mathematics, 2009.

\end{thebibliography}\endgroup
\bibliographystyle{utphys}

\end{document}